\documentclass[preprint,12pt]{aastex}
\newcommand{\fracb}[2]{\left(\frac{#1}{#2}\right)}
\newcommand{\fracsb}[2]{\left[\frac{#1}{#2}\right]}
\newcommand{\mean}[1]{\langle{#1}\rangle}

\begin{document}

\title{On the lateral expansion of GRB jets}

\author{Jonathan Granot\altaffilmark{1,2,3} and Tsvi Piran\altaffilmark{1}}

\altaffiltext{1}{Racah Institute of Physics, The Hebrew University, Jerusalem 91904, Israel}
\altaffiltext{2}{Raymond and Beverly Sackler School of Physics \& Astronomy, Tel Aviv University, Tel Aviv 69978, Israel}
\altaffiltext{3}{Centre for Astrophysics Research, University of Hertfordshire, College Lane, Hatfield, AL10 9AB, UK}

\begin{abstract}

The dynamics of GRB jets during the afterglow phase have an important
effect on the interpretation of their observations and for inferring
key physical parameters such as their true energy and event
rate. Semi-analytic models generally predict a fast lateral expansion,
where the jet opening angle asymptotically grows exponentially with
its radius. Numerical simulations, however, show a much more modest
lateral expansion, where the jet retains memory of its initial opening
angle for a very long time, and the flow remains non-spherical until
it becomes sub-relativistic, and only then gradually approaches
spherical symmetry. Here we suggest a new analytic model based on a
new physically derived recipe for the lateral expansion.  We also
generalize the model by relaxing the common approximations of
ultra-relativistic motion and a narrow jet opening angle.  We find
that the new analytic model fits much better the results of numerical
simulations, mainly because it remains valid also in the mildly
relativistic, quasi spherical regime.  This model shows that for
modest initial jet half-opening angles, $\theta_0$, the outflow is not
{\it sufficiently} ultra-relativistic when its Lorentz factor reaches
$\Gamma = 1/\theta_0$ and therefore the sideways expansion is rather
slow, showing no rapid, exponential phase.  On the other hand, we find
that jets with an extremely narrow initial half-opening angle, of
about $\theta_0 \ll 10^{-1.5}$ or so, which are still sufficiently
ultra-relativistic at $\Gamma = 1/\theta_0$, do show a phase of rapid,
exponential lateral expansion.  However, even such jets that expand
sideways exponentially are still not spherical when they become
sub-relativistic.

\end{abstract}

\maketitle

\section{Introduction}

The ultra-relativistic outflows that power gamma-ray bursts (GRBs) are
thought to be collimated into narrow jets \citep[for reviews
see][]{Piran05,Granot07,GR-R11}. The evidence for this is rather
indirect, however, since their images are usually unresolved, and in
the best case (GRB~030329) the late time radio afterglow image was
only marginally
resolved~\citep{Frail97,Taylor97,Taylor04,Pihlstrom07}. The different
lines of evidence for jets in GRBs include analogy to other
astrophysical relativistic outflow sources such as active galactic
nuclei or micro-quasars~\citep[e.g.][]{Rhoads97}, the difficulty in
transferring enough energy to ultra-relativistic ejecta in a spherical
explosion of a massive star~\citep[for long duration
GRBs;][]{TMM01,PV02,Granot07}, extremely large isotropic equivalent
energies in some GRBs \citep[with $E_{\rm\gamma,iso}\approx 4.9M_\odot
c^2$ in GRB080916C][]{GRB080916C}, and an achromatic steepening of the
afterglow lightcurves of some GRBs that is attributed to a
jet~\citep[known as a ``jet
break'';][]{Rhoads97,Rhoads99,SPH99,Fruchter99,Harrison99,Kulkarni99,Halpern00,Price01}.
Therefore, there is very little direct observational information about
the jet angular structure and dynamics, which make it difficult to
interpret GRB afterglow observations and infer from them important
physical parameters such as the jet energy and opening angle, the
external density profile, and the microphysical parameters of the
relativistic collisionless shock powering the afterglow emission.

Most studies of GRB jet dynamics during the afterglow phase have
focused on a roughly uniform jet with well defined, sharp edges. We
shall also focus on such a uniform jet, and only briefly remark on the
expected relation to jets with a smoother angular structure (also
known as ``structured jets''). The jet dynamics have been studied both
analytically~\citep{Rhoads99,SPH99,PM99,KP00,MSB00,Piran00,ONP04,Granot07}
and numerically, using two dimensional special relativistic numerical
simulations~\citep{Granot01,CGV04,ZM09,vanEerten10,MK10,WWF11,vEM11},
as well as an intermediate approach where the dynamical equations are
integrated over the radial profile of the thin shocked region, thus
reducing the set of partial differential equations to one
dimension~\citep{KG03}.

Let us consider a uniform double-sided jet of total energy $E_{\rm
jet}$, initial half-opening angle $\theta_0$, and initial Lorentz
factor $\Gamma_0$.  GRB observations suggest that typically
$\Gamma_0\theta_0\gg 1$. At early times, as long as
$\Gamma\gg\theta_0^{-1}$, the bulk of the jet is causally disconnected
from its edge and thus evolves as if it where part of a spherical flow
with an energy $E_{\rm iso} = (1-\cos\theta_0)^{-1}E_{\rm jet} \approx
2\theta_0^{-2}E_{\rm jet}$, following the spherical~\citet{BM76}
self-similar solution. This early phase corresponds to radii $R <
R_j$, where the jet radius $R_j$ is defined as the radius where
$\Gamma = 1/\theta_0$ for a spherical flow with $E = E_{\rm iso}$. At
$R>R_j$ the bulk of the jet is in causal contact with its
edge and the jet can in principal rapidly expand sideways. However,
the degree of lateral spreading at this stage, which strongly affects
the dynamics, is not well known. Therefore, the jet dynamics at
$R>R_j$ are still controversial. In particular, the radius $R_{\rm
NR}$ at which the flow (or jet) becomes non-relativistic, still
remains uncertain.

The Sedov length for a spherical flow with the true jet energy, $E =
E_{\rm jet}$ (i.e. the radius where it sweeps up a rest mass energy
equal to its own energy and becomes non-relativistic), $R_{\rm
S}(E_{\rm jet})$, is very close to $R_j$.  Therefore, in order for the
jet to be already close to spherical when it becomes non-relativistic
(i.e. at $R_{\rm NR}$), it must expand sideways very quickly and
become close to spherical already near $R_j$ [i.e. $R_{\rm NR}$ cannot
be $\gg R_j\sim R_{\rm S}(E_{\rm jet})$]. This is indeed roughly what
happens in simple analytic models, where the jet half-opening angle,
$\theta_j$, starts growing exponentially with radius near $R_j$, and
the jet quickly becomes close to spherical and non-relativistic at a
radius $R_{\rm NR}\sim (1-\ln\theta_0)R_j\sim (1-\ln\theta_0)R_{\rm
S}(E_{\rm jet})$, which is larger than $R_j$ only by a logarithmic
factor\footnote{The mild discrepancy, by a logarithmic factor, between
$R_{\rm NR}$ and $R_{\rm S}(E_{\rm jet})\sim R_j$ likely arises from
the fact that in simple analytic models the swept-up mass at the
radius where the jet becomes spherical is smaller than the external
rest mass within a sphere of the same radius.}, while at $R>R_{\rm
NR}$ the flow quickly approaches the Newtonian, spherical,
self-similar Sedov-Taylor solution.

Numerical simulations, however, suggest that most of the energy
remains within the initial jet half-opening angle $\theta_0$ until the
flow becomes mildly relativistic, and only then does the flow start to
gradually approach spherical
symmetry~\citep{Granot01,CGV04,ZM09,vanEerten10,MK10}. Under
the crude approximation that the jet does not expand sideways and
keeps evolving as a conical section of a spherical flow up until the
radius where it becomes non-relativistic, the latter is given by
$R_{\rm NR} \sim R_{\rm S}(E_{\rm iso}) =\theta_0^{-2/(3-k)}R_j$.  In
this case the flow is still highly non-spherical at $R_{\rm NR}$, and
only very gradually approaches spherical symmetry~\citep{GR-RL05}.

This clearly shows that without lateral expansion $R_{\rm NR}$ is
significantly larger, by a factor of
$\sim\theta_0^{-2/(3-k)}/(1-\ln\theta_0)$ (which is $\gg 1$ for
$\theta_0\ll 1$), than if there is fast lateral expansion at $R>R_j$.
Thus, the dynamics of the flow at small radii ($R\ll R_j$) and at
large radii ($R\gg R_{\rm S}(E_{\rm iso})$) are reasonably well known,
while at intermediate radii ($R_j\lesssim R\lesssim R_{\rm S}(E_{\rm
iso})$) they are still controversial. For typical values of $\theta_0
\sim 0.1$ this range of radii may appear rather small, $R_{\rm
NR}/R_j\sim 1-\ln\theta_0 \sim 3.3$ for exponential lateral expansion
and $R_{\rm NR}/R_j\sim\theta_0^{-2/(3-k)} \sim 4.6$ for no lateral
expansion up to $R_{\rm NR} \sim R_{\rm S}(E_{\rm iso})$ with $k =
0$. However, it corresponds to a large range in observed times (over
which the corresponding afterglow emission reaches us), of $t_{\rm
obs,NR}/t_{{\rm obs},j}\sim (1-\ln\theta_0)\theta_0^{-2} \sim 330$ and
$\sim\theta_0^{-(8-2k)/(3-k)} \sim 460$, respectively, since the
observed time scales as $t_{\rm obs} \sim R/c\Gamma^2$, and $\Gamma$
decreases by a large factor (of $\theta_0^{-1}$) within this range of
radii.

Most simulations so far were for $\theta_0 = 0.2$, or even wider
initial jet half-opening angles. Recently, however, \citet{WWF11} and
later \citet{vEM11} have performed simulations also for narrower
initial jets, $\theta_0 = 0.05,\,0.1,\,0.2$. \citet{WWF11} have found
that significant lateral spreading starts when $\Gamma$ drops below
$\theta_0^{-1}$, as predicted by analytic models, and tried to
reconcile the apparent discrepancy with analytic models by attributing
it to their small range of validity after significant lateral
spreading starts ($1\ll\Gamma<\theta_0^{-1}$) for the typical modest
values of $\theta_0$ used in the simulations. \citet{vEM11} disagree
with this conclusion, and we address this dispute in \S~\ref{sec:dis}.
More recently, \citet{Lyut11} has argued that significant lateral
spreading is expected only at a later stage, when $\Gamma$ drops below
$\theta_0^{-1/2}$ (rather than $\theta_0^{-1}$ as obtained in simple
analytic models), based on an analytic consideration (which we address
in \S~\ref{sec:recipe} and Appendix~\ref{sec:comp-prev}, and find to
be in error). Thus, there appears to be an ongoing debate on these
important issues.
     
Here we try to reconcile the apparent differences between the analytic
and numerical results, in light of this recent debate. The different
relevant critical radii are discussed in
\S~\ref{sec:R}. In \S~\ref{sec:recipe} we discuss the recipe for
lateral expansion used by analytic models, and derive a new recipe
that takes into account the non-spherical nature of the shock driven
by the jet into the external medium. In \S~\ref{sec:model} we
construct an analytic relativistic model, which includes both the
traditional recipe and our new recipe for the jet lateral
expansion. It is also shown that while the region of interest and
validity of the analytic model (corresponding to
$1\ll\Gamma<\theta_0^{-1}$) increases as $\theta_0$
decreases, $\theta_j$ reaches lower values, resulting in a narrower
jet at the time when the analytic solution becomes invalid.  Because
this relativistic model breaks down in a region of interest (both for
typical GRB parameters and for comparison with simulations), in
\S~\ref{sec:gen} we generalize it so that it would be valid also at 
low $\Gamma$ and high $\theta_j$, using two different assumptions on
the accumulation of the swept-up external mass (in
\S~\ref{sec:trumpet} and \S~\ref{sec:conical}). According to the
results of these models (\S~\ref{sec:model-results}), a phase of rapid
exponential lateral expansion exists only for sufficiently narrow
initial jet half-opening angles, of approximately $\theta_0\ll
0.05,\,0.03,\,0.01$ for $k = 0,\,1,\,2$, respectively.  In
\S~\ref{sec:comp-sim} we compare our analytic models to numerical
simulations (with a modest $\theta_0 = 0.2$) and find reasonably good
agreement (with no exponential lateral expansion in both cases), where
the differences between the two recipes for the lateral spreading have
a smaller effect on the agreement with numerical simulations compared
to the generalization of the model to small $\Gamma$ and large
$\theta_j$.  The implications of our results are discussed in
\S~\ref{sec:dis}.

\section{The different critical radii, and two extreme assumptions for the jet dynamics}
\label{sec:R}

Using the approximate equation for energy conservation (for $\Gamma\gg
1$), $E\approx\Gamma^2M(R)c^2$, where $M(R)$ is the swept-up rest mass
at radius $R$ for a spherical flow in an external density $\rho_{\rm
ext} = A R^{-k}$ (with $k < 3$), and the definition of the jet
radius $R_j$ as the radius where $\Gamma=\theta_0^{-1}$ for a
spherical flow of energy $E_{\rm iso}$, we obtain
\begin{equation}
R_j = \fracsb{(3-k)E_{\rm jet}}{2\pi Ac^2}^{1/(3-k)}
= 2^{1/(3-k)}R_{\rm S}(E_{\rm jet})\ .
\end{equation}
Similarly, the Sedov radius of a spherical flow with $E=E_{\rm iso}$ is
\begin{equation}
R_{\rm S}(E_{\rm iso}) = \fracsb{(3-k)E_{\rm iso}}{4\pi Ac^2}^{1/(3-k)}
= \theta_0^{-2/(3-k)}R_j = \fracb{\theta_0^2}{2}^{-1/(3-k)}R_{\rm S}(E_{\rm jet})\ .
\end{equation}

Two extreme assumptions on the degree of lateral spreading, which
likely bracket the true jet dynamics are: 1. mildly relativistic
lateral expansion in the jet co-moving frame, and 2. no lateral
spreading until the jet becomes non-relativistic. Assumption 1, which
is made in most semi-analytic models, results in exponential growth of
$\theta_j(R)$, until the jet becomes quasi-spherical and
non-relativistic at 
\begin{equation}
R_{\rm NR,1}\sim(1-\ln\theta_0)R_j\quad\quad{\rm (fast\ lateral\ spreading)}\ . 
\end{equation}
Assumption 2 was so far studied mainly by \citet{GR-RL05}, and leads
to
\begin{equation}
R_{\rm NR,2} = R_{\rm S}(E_{\rm iso}) 
= \theta_0^{-2/(3-k)}R_j\quad\quad{\rm (no\ lateral\ spreading)}\ .
\end{equation}
In this case the jet is still very far from spherical symmetry at
$R_{\rm NR}$, and thus approaches spherical symmetry only after the
radius grows by a factor $b_2$, of a few or several. Moreover, since
the radius of the Sedov-Taylor solution scales as
\begin{equation}
R_{\rm ST}(E,t) \sim R_{\rm S}(E)\fracsb{ct}{R_{\rm S}(E)}^{2/(5-k)} 
\sim \fracb{Et^2}{A}^{1/(5-k)}\ ,
\end{equation}
and at the non-relativistic transition time, $t_{\rm NR,2}\sim R_{\rm
NR,2}/c  = R_{\rm S}(E_{\rm iso})/c$, the Sedov-Taylor radius of a
spherical flow with the true jet energy is much smaller than the jet
radius at that time, $R_{\rm ST}(E_{\rm jet},t_{\rm NR,2})/R_{\rm NR,2}
\sim \theta_0^{-2/(5-k)}\ll 1$, the flow approaches spherical symmetry 
only at the time $t_{\rm sph,2}$ when $R_{\rm ST}(E_{\rm jet},t_{\rm
sph,2}) = b_2R_{\rm NR,2} = b_2R_{\rm S}(E_{\rm iso})$, which
corresponds to $t_{\rm sph,2}/t_{\rm NR,2} \sim
\theta_0^{-1}b_2^{(5-k)/2}\gg 1$~\citep[see Eq. 6 of][]{GR-RL05}. Note
that this is much smaller than the factor ($b_2$) by which the radius
grows over the same time.  For assumption 1 similar arguments imply
$t_{\rm sph,1}/t_{\rm
NR,1}\sim(1-\ln\theta_0)^{(3-k)/2}b_1^{(5-k)/2}$, where $b_1 < b_2$
can be expected.

\begin{figure}[!t]
\begin{center}
\includegraphics[width=0.90\columnwidth]{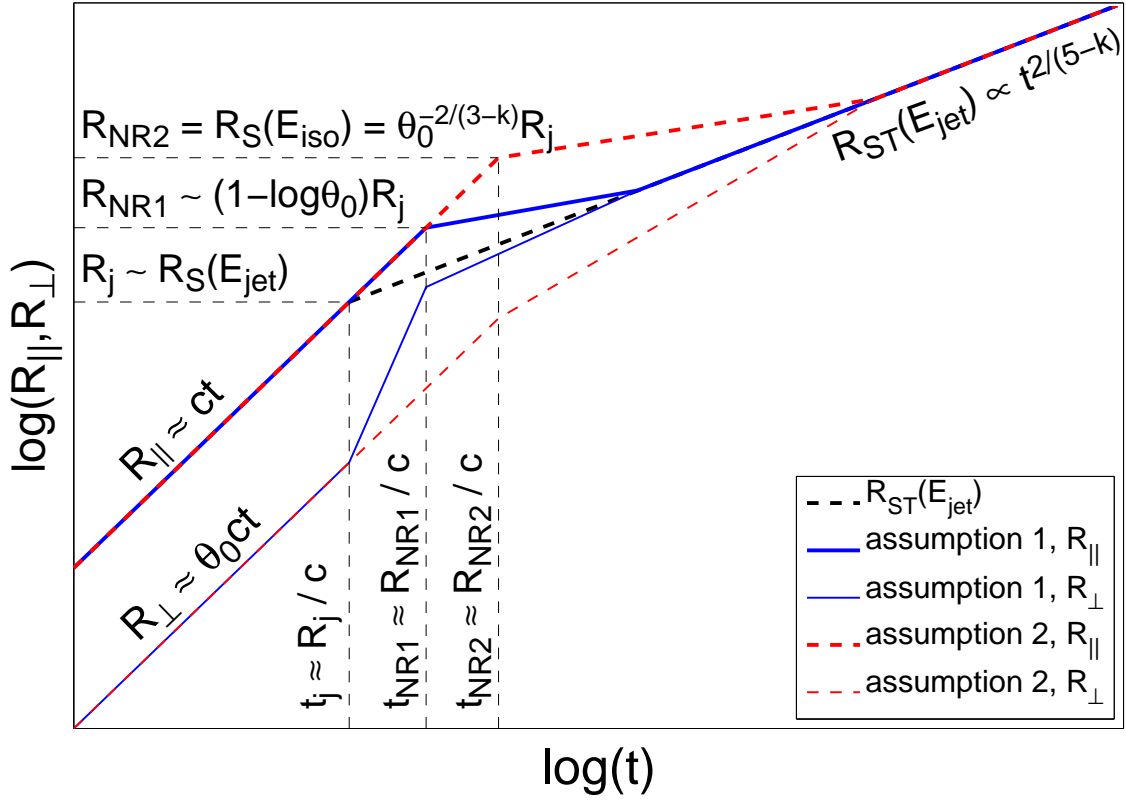}
\caption{A schematic figure showing the evolution of the jet radius 
$R = R_\parallel$ (i.e. its extent along its symmetry axis) and
lateral size $R_\perp$ as a function of the lab frame time $t$ for two
extreme assumptions on its degree of lateral spreading: (1) mildly
relativistic lateral expansion in the jet co-moving frame, and (2) no
lateral spreading until the jet becomes non-relativistic. The jet
becomes spherical when $R_\parallel$ and $R_\perp$ become equal, which
occurs well after the jet becomes non-relativistic, and then joins the
Sedov-Taylor solution.
\label{fig:R1}}
\end{center}
\end{figure}

Fig.~\ref{fig:R1} shows the jet radius $R = R_\parallel$ (i.e. its
extent along its symmetry axis) and lateral size $R_\perp$ as a
function of the lab frame time $t$ for these two assumptions.  The
region where the dynamics for these two extreme assumptions differ is
basically where the dynamics are most uncertain, and corresponds to
the range of radii $R_{\rm S}(E_{\rm jet}) < R < b_2R_{\rm S}(E_{\rm
iso})$ (i.e. a factor of $f_R\sim
\theta_0^{-2/(3-k)}b_2$ in radius), and (lab frame) times 
$R_{\rm S}(E_{\rm jet})/c < t < t_{\rm sph,2}\sim
\theta_0^{-1}b_2^{(5-k)/2}R_{\rm S}(E_{\rm iso})/c$ (or a factor of $f_t
\sim f_R^{(5-k)/2}\sim\theta_{0}^{-(5-k)/(3-k)}b_2^{(5-k)/2}$ in time).

Altogether, the ordering of the different radii is
\begin{equation}
R_{\rm S}(E_{\rm jet}) \sim R_j < R_{\rm NR,1} < R_{\rm S}(E_{\rm iso}) = R_{\rm NR,2}\ ,
\end{equation}
or
\begin{equation}
1 \sim \frac{R_{\rm S}(E_{\rm jet})}{R_j} \sim (1-\ln\theta_0)^{-1}\frac{R_{\rm NR,1}}{R_j} 
\sim \theta_0^{-2/(3-k)}\frac{R_{\rm S}(E_{\rm iso})}{R_j} 
= \theta_0^{-2/(3-k)}\frac{R_{\rm NR,2}}{R_j}\ .
\end{equation}

\section{Analytic recipe for lateral expansion}
\label{sec:recipe}

The ``traditional'' basic underlying model assumptions used for the
analytic modeling of relativistic jet dynamics during the afterglow
phase~\citep[e.g.,][]{Rhoads99,SPH99} are (i) a uniform jet within a
finite half-opening angle $\theta_j$ with an initial value $\theta_0$
that has sharp edges, (ii) the shock front is part of a sphere at any
given lab frame time $t$, (iii) the outer edge of the jet is expanding
sideways mildly relativistically, with $u'_\theta \sim 1$ in the local
rest frame of the jet (where quantities are denoted with a prime),
(iv) the jet velocity is always in the radial direction and $\theta_j
\ll 1$. Under these assumptions, the jet dynamics are obtained by
solving the 1D ordinary differential equations for the conservation of
energy and particle number.\footnote{For the adiabatic energy
conserving evolution considered here, the equation for momentum
conservation is trivial in spherical geometry, and does not constrain
the dynamics. For a narrow ($\theta_j \ll 1$) highly relativistic
($\Gamma\gg 1$) jet, the equation for the conservation of linear
momentum in the direction of the jet symmetry axis is almost identical
to the energy conservation equation. When the jet becomes
sub-relativistic the conservation of energy and linear momentum force
it to approach spherical symmetry, and once it becomes quasi-spherical
then again the momentum conservation equation becomes irrelevant.}

The lateral expansion speed in the lab frame (i.e. the rest frame of
the central source and the external medium) is $\beta_\theta =
u_\theta/\Gamma = u'_\theta/\Gamma$, where $u_\theta
=\Gamma\beta_\theta$ is its lateral component of the 4-velocity (which
is Lorentz invariant, so that $u'_\theta = u_\theta$), while $u_r
=\Gamma\beta_r$ is its radial component. Primed quantities are
measured in a frame moving at $\beta_r\hat{r}$ in the radial
direction, so that $\beta'_r = 0$ and $\beta' =
[1-(\Gamma')^{-2}]^{1/2} = \beta'_\theta$.  The usual
assumption~\citep{Rhoads99,SPH99} is that $u'_\theta \sim 1$, which
corresponds to
\begin{equation}\label{eq:old-recipe}
\beta_\theta \sim \frac{1}{\Gamma}\ .
\end{equation}
As is shown in the next section, $\beta_\theta\approx d\theta_j/d\ln
R$ directly determines the jet lateral expansion rate in the lab
frame.

Here we derive a new physically motivated recipe. It relies on the
fact that for any shock front, with an arbitrary shape, the local
velocity vector of the material just behind the shock front as
measured in the rest frame of the upstream fluid ahead of the shock
(i.e. the lab frame in our case), $\vec{\beta}$, is normal to the
shock front~\citep[i.e. in the direction of the shock normal,
$\hat{n}$, at that location;][]{KG03}, namely
\begin{equation}\label{n_beta_hat}
\hat{\beta} = \hat{n}\ .
\end{equation}
A simple way to understand this result is that as each fluid element
passes through the shock it samples only the local conditions (and is
not aware of the large scale or global shock front geometry) and
locally the shock normal is the only preferred direction in the
upstream rest frame (e.g., the pressure gradients that accelerate the
fluid element are in the $-\hat{n}$ direction and thus accelerate it
in the $\hat{n}$ direction). For an axisymmetric shock (with no
dependence on the azimuthal angle $\phi$), Eq.~(\ref{n_beta_hat})
immediately implies that the angle $\alpha$ between the shock normal,
$\hat{n}$, and the radial direction, $\hat{r}$, which is defined by
$\cos\alpha\equiv\hat{n}\cdot\hat{r} =\hat{\beta}\cdot\hat{r}$,
satisfies
\begin{equation}\label{eq:alpha}
\tan\alpha = \frac{\beta_\theta}{\beta_r} =
-\frac{1}{R}\frac{\partial R}{\partial\theta}
= -\frac{\partial\ln R}{\partial\theta}\ ,
\end{equation}
where $\theta$ is the polar angle measured from the jet symmetry axis.
Since $R\sim \beta ct$, we have $\partial\ln
R/\partial\theta\sim\partial\ln\beta/\partial\theta =
\Gamma^{-2}\partial\ln u/\partial\theta \sim
-1/\Gamma^2\Delta\theta$, where $\Delta\theta$ is the angular scale
over which $u$ varies significantly, and we have assumed that
$u$ decreases with $\theta$, as is usually expected. Since for
$\Gamma\gg 1$ and $\alpha\ll 1$ we also have $\beta_r\approx 1$,
Eq.~(\ref{eq:alpha}) implies that $\beta_\theta \sim
1/\Gamma^2\Delta\theta$. For a roughly uniform jet of half-opening
angle $\theta_j$ we have $\Delta\theta \sim \theta_j$, and therefore
\begin{equation}\label{eq:new-recipe}
\beta_\theta \sim \frac{1}{\Gamma^2\Delta\theta} \sim \frac{1}{\Gamma^2\theta_j}\ ,
\end{equation}
which is our new recipe for lateral expansion.

Eq. (\ref{eq:new-recipe}) was first derived in the context of GRBs by
\cite{KG03}. Recently it was rederived by \cite{Lyut11}, based on an
earlier work by \cite{Shapiro79}. \citet{Lyut11} has argued that
Eq. (\ref{eq:new-recipe}) implies a negligible lateral expansion as
long as $\Gamma > 1/ \sqrt{\theta_j}$ suggesting that with this model
one obtains a slow sideways expansion, as seen in the numerical
simulations.  However, as we show later, this formula results in a
slower lateral expansion (compared to the usual recipe,
i.e. Eq.~[\ref{eq:old-recipe}]) only as long $\Gamma > 1/\theta_j$
(the standard condition for the onset of significant lateral
expansion), but once $\Gamma < 1/\theta_j$ this formula leads to a
faster sideways expansion. We also show later that other factors,
namely the break down of the ultra-relativistic and small angle
approximations, are the main cause for the discrepancy between the
existing simple analytic models and the numerical simulations.  For
completeness we discuss the details of Lyutikov's and Shapiro's work
in Appendix~A.

\section{A simple relativistic model}
\label{sec:model}

We turn now to compare the traditional recipe for the lateral
expansion speed, $\beta_\theta\sim 1/\Gamma$
(Eq.~[\ref{eq:old-recipe}]), with our own new simple recipe,
$\beta_\theta \sim 1/\Gamma^2\theta_j$ (Eq.~[\ref{eq:new-recipe}]),
which was derived in the previous section.  These recipes are
implemented here within the semi-analytic model for the jet dynamics
of~\citet{Granot07}. The main results are provided here and we refer
the reader to that work for more details on that model. Broadly
similar semi-analytic models, with some variations, were used earlier
by other
authors~\citep[e.g.,][]{Rhoads97,Rhoads99,SPH99,PM99,KP00,MSB00,ONP04}.

The lateral size of the jet, $R_\perp$, and its radius, $R =
R_\parallel$, are related by $R_\perp \approx \theta_j R$. The
evolution of $R_\perp$ is governed by
\begin{equation}\label{dR_perp}
dR_\perp \approx \theta_j dR + \beta_\theta cdt \approx 
\left(\theta_j + \beta_\theta\right)dR\ ,
\end{equation}
and therefore 
\begin{equation}\label{dtheta_dR}
\frac{d\theta_j}{d\ln R} \approx \beta_\theta 
\approx \frac{1}{\Gamma^{1+a}\theta_j^a}\ ,\quad\quad
a = \left\{\matrix{
1\quad\quad (\hat{\beta}=\hat{n})\ , \cr \cr
0\quad (u'_\theta\sim 1)\ ,}\right.
\end{equation}
where we have conveniently introduced the parameter $a$ that enables us
to analyze these two different recipes together.

The external density is assumed to be a power law in
radius\footnote{We consider here and throughout this review only $k <
3$ for which the shock Lorentz factor decreases with radius for a
spherical adiabatic blast wave during the self-similar stage of its
evolution \citep{BM76}.}, $\rho_{\rm ext} = AR^{-k}$.  The total
swept-up (rest) mass, $M(R)$, is accumulated as
\begin{equation}\label{dM_dR}
\frac{dM}{dR} \approx 2\pi(\theta_jR)^2\rho_{\rm ext}(R) 
=  2\pi AR^{2-k}\theta_j^2(R)\ ,
\end{equation}
where the factor of $2$ is since a double sided jet is assumed. As
long as the jet is relativistic, energy conservation takes the form
$E_{\rm jet}\approx \Gamma^2 Mc^2$, which implies that $Md(\Gamma^2) =
-\Gamma^2dM$, and
\begin{equation}\label{dGamma_dR}
\frac{d\Gamma}{dR} = -\frac{\Gamma}{2M}\frac{dM}{dR} = -\pi A
R^{2-k}\theta_j^2(R)\frac{\Gamma(R)}{M(R)}\ .
\end{equation}
One can numerically integrate equations (\ref{dtheta_dR}),
(\ref{dM_dR}), and (\ref{dGamma_dR}) thus obtaining $\theta_j(R)$,
$M(R)$, and $\Gamma(R)$. Alternatively, one can use the relation
$E_{\rm jet}\approx \Gamma^2 Mc^2$ (energy conservation) which reduces
the number of free variable to two, and solve equations
(\ref{dtheta_dR}) and (\ref{dGamma_dR}). Changing to normalized
dimensionless variables $\theta\equiv \theta_j/\theta_0$,
$\gamma\equiv\Gamma\theta_0$ and $r\equiv[(3-k)/2]^{1/(3-k)}R/R_j$,
gives
\begin{eqnarray}\label{dtheta_dRt}
\frac{d\theta}{dr} &=&
r^{-1}\,\gamma^{-1-a}(r)\,\theta^{-1}(r)\ ,
\\ \label{dGamma_dRt}
\frac{d\gamma}{dr} &=& 
-\,r^{2-k}\,\gamma^3(r)\,\theta^2(r)\ ,
\end{eqnarray}
where the initial conditions at some small radius $r_0 \ll 1$ (just
after the deceleration radius) are 
\begin{equation}\label{IC}
\theta(r_0) = 1\ ,\quad\quad
\gamma(r_0) = 
\sqrt{\frac{3-k}{2}}\,r_0^{-(3-k)/2}\ .
\end{equation}

Note that by definition, $\gamma\theta =
\Gamma\theta_j$. Eqs. (\ref{dtheta_dRt}) and (\ref{dGamma_dRt}) imply
\begin{equation}\label{dy_j_dRt}
\frac{d(\gamma\theta)}{dr} = \frac{1}{r(\gamma\theta)^a}
-r^{2-k}(\gamma\theta)^3 =
\frac{1-r^{3-k}(\gamma\theta)^{3+a}}{r(\gamma\theta)^a}\ .
\end{equation}
For $r\ll 1$ the second term on the r.h.s of Eq.~(\ref{dy_j_dRt})
dominates, implying $(\gamma\theta)^2 \approx
\frac{3-k}{2}r^{k-3}$, which is consistent with Eq.~(\ref{IC}).
This suggests that the two terms become comparable at $r \approx
r_c$ that is given by
\begin{equation}\label{eq:r_c}
r_c = \fracb{3-k}{2}^{(3+a)/[(1+a)(3-k)]}\ . 
\end{equation}
While $r_c > 1$ for $k<1$, it can reach very low values ($r_c\ll 1$)
as $k$ approaches $3$. We are interested here mainly in $k \geq 2$,
for which $r_c\sim 1$ still approximately holds. We do note, however,
that the lower values of $r_c$ for higher values of $k$ result in an
earlier onset of significant lateral expansion for such higher
$k$-values. Now let us examine what happens at $r\gg r_c \sim 1$. If
we assume that the first term becomes dominant then
Eq.~(\ref{dy_j_dRt}) would imply $\gamma\theta \approx [(1+a)\ln
r]^{1/(1+a)}$, which in turn implies that the second term would be
dominant (since $(\gamma\theta)^{3+a}r^{3-k}\approx [(1+a)\ln
r]^{(3+a)/(1+a)}r^{3-k}\gg 1$), rendering the original assumption
inconsistent.  The same applies if the opposite assumption is made,
that the second term is dominant (in this case
$\gamma\theta\approx\sqrt{\frac{3-k}{2}}r^{(k-3)/2}$ which implies
that the first term would be dominant,
$(\gamma\theta)^{3+a}r^{3-k}\approx\fracb{3-k}{2}^{(3+a)/2}r^{(k-3)(1+a)/2}\ll
1$).  This implies that the two terms must remain comparable, implying
$\gamma\theta\sim r^{(k-3)/(3+a)}$.  A similar conclusion can be reach
by taking the ratio of equations (\ref{dtheta_dRt}) and
(\ref{dGamma_dRt}) which implies that
\begin{equation}
d(\theta^{3+a}) = r^{k-3}d(\gamma^{-3-a})\ .
\end{equation}
A more careful examination shows that they must cancel each other to
leading order, and the first two leading terms for $r\gg 1$
are given by
\begin{equation}\label{y_j}
\gamma\theta \approx r^{(k-3)/(3+a)}+\frac{3-k}{(3+a)^2}\,r^{(k-3)(2+a)/(3+a)}\ .
\end{equation}
Substituting equation (\ref{y_j}) into equations (\ref{dtheta_dRt})
and (\ref{dGamma_dRt}) yields
\begin{eqnarray}
\frac{d\ln\theta}{d\ln r} &\approx &
r^{(3-k)(1+a)/(3+a)}-\frac{(3-k)(1+a)}{(3+a)^2}\ ,
\\
\frac{d\ln\gamma}{d\ln r} &\approx & 
-r^{(3-k)(1+a)/(3+a)}-\frac{2(3-k)}{(3+a)^2}\ ,
\end{eqnarray}
and
\begin{eqnarray}\label{theta_scaling}
\theta &\approx& b\,r^{-\frac{(3-k)(1+a)}{(3+a)^2}}
\exp\left[\frac{(3+a)}{(3-k)(1+a)}\,r^\frac{(3-k)(1+a)}{(3+a)}\right]\ ,\quad 
\\ \label{Gamma_scaling} 
\gamma &\approx& \frac{1}{b}\,r^{-\frac{2(3-k)}{(3+a)^2}}
\exp\left[-\frac{(3+a)}{(3-k)(1+a)}\,r^\frac{(3-k)(1+a)}{(3+a)}\right]\ ,
\quad\quad
\end{eqnarray}
where the normalization coefficient $b$ is determined numerically.
For $r\ll 1$ we have
\begin{equation}\label{y_j1}
\gamma\theta \approx \sqrt{\frac{3-k}{2}}\,r^\frac{k-3}{2}+
\fracb{3-k}{2}^\frac{-2-a}{2}\frac{r^\frac{a(3-k)}{2}}{(3+a)}\ .
\end{equation}

\begin{figure}
\begin{center}
\includegraphics[width=0.80\columnwidth]{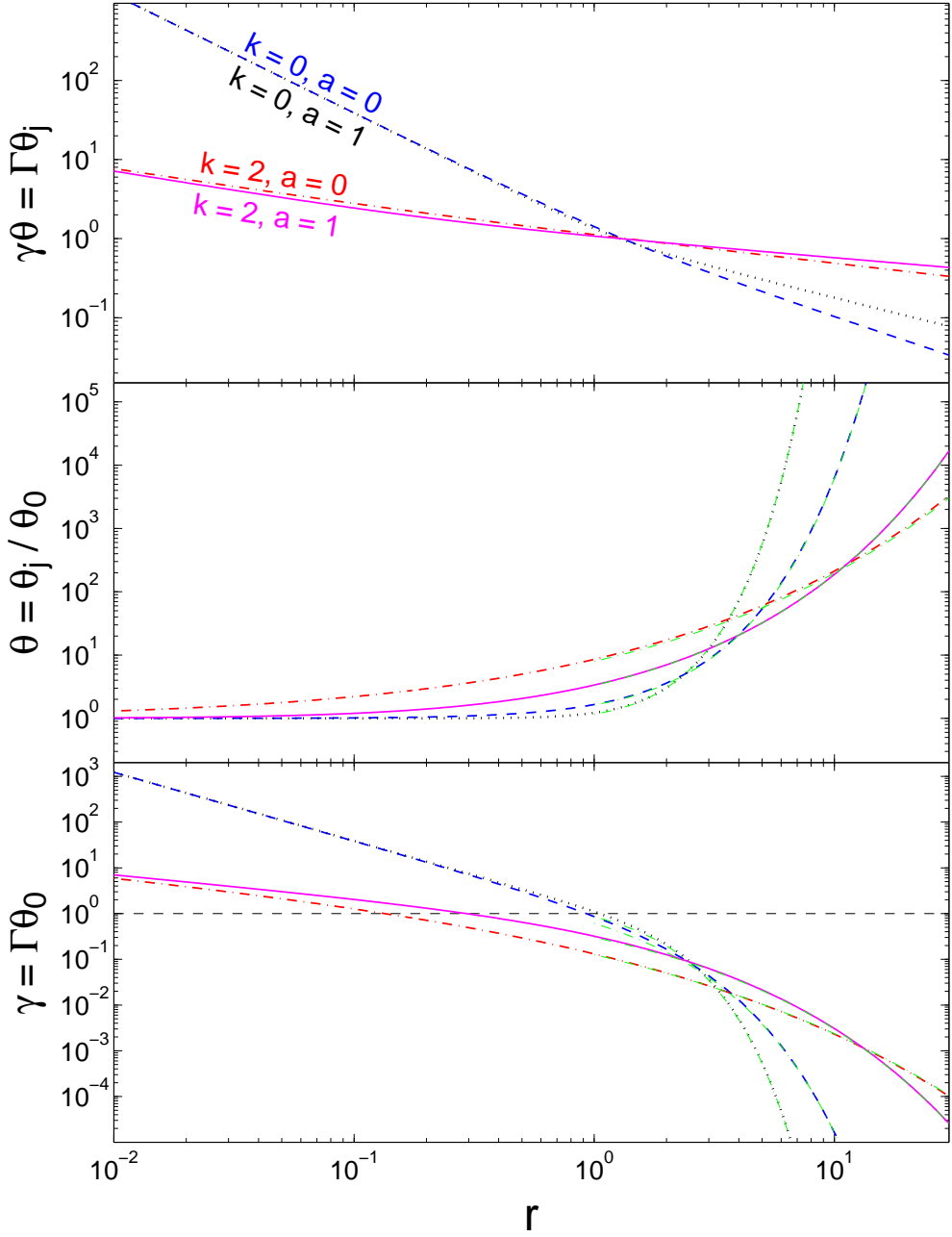}
\caption{The jet dynamics according to our relativistic analytic model
  (see text in \S~\ref{sec:model} for details), for either a uniform
  ($k=0$) or a wind-like, stratified ($k=2$) external density profile,
  and for either the old ($a=0$) or our new ($a=1$) recipe for the jet
  lateral expansion speed. The dynamical range in this figure is
  unrealistically large, and it is shown mainly in order to
  demonstrate the properties of this solution, and show how well our
  analytic approximation for $r > 1$ works (the {\it dashed green
  lines} in the middle and bottom panels, which are practically on top
  of the numerical results).
\label{fig:jet_dyn_1D}}
\end{center}
\end{figure}

Fig.~\ref{fig:jet_dyn_1D} shows the results of our model in terms of
the normalized jet half-opening angle $\theta =
\theta_j/\theta_0$ and Lorentz factor $\gamma =
\Gamma\theta_0$ (as well as their product,
$\gamma\theta = \Gamma\theta_j$) as a function of the normalized
radius $r = [(3-k)/2]^{1/(3-k)}R/R_j$. The results are shown both for
a uniform external medium ($k=0$), which is the main focus of this
work, as well as for a stellar wind ($k=2$; this is included mainly
for completeness and is only briefly discussed in \S~\ref{sec:dis}).
The dynamical range in this figure is unrealistically large, and it is
shown mainly in order to demonstrate the properties of this solution,
and show how well our analytic approximation for $r > 1$ works (the
dashed green lines in the middle and bottom panels, which are
practically on top of the numerical results). The excellent agreement
between our semi-analytic results (the numerical solution of
Eqs.~[\ref{dtheta_dRt}] and [\ref{dGamma_dRt}]) and analytic formulas
(Eqs.~[\ref{theta_scaling}] and [\ref{Gamma_scaling}]) shows that our
analytic results (including Eqs.~[\ref{y_j}] and [\ref{y_j1}]) can be
safely used in order to analyze the result of this model. This good
agreement was also used in order to find the exact values of the
numerical coefficient $b$ that determines the normalization for
$\theta$ and $\gamma$., which were found to be $b(k=0,a=0)\approx
b(k=0,a=1)\approx 0.60$, $b(k=2,a=0)\approx 0.395$ and
$b(k=2,a=1)\approx 0.45$.  Our new recipe for the lateral expansion
speed (Eq.~[\ref{eq:new-recipe}]) results in a slower initial lateral
expansion compared to the old recipe at $r \ll 1$, where
$\Gamma\theta_j = \gamma\theta \gg 1$. However, at larger radii,
$r\gtrsim 1$, where $\Gamma\theta_j < 1$ it results in a faster
lateral expansion.

\begin{figure}
\begin{center}
\includegraphics[width=0.85\columnwidth]{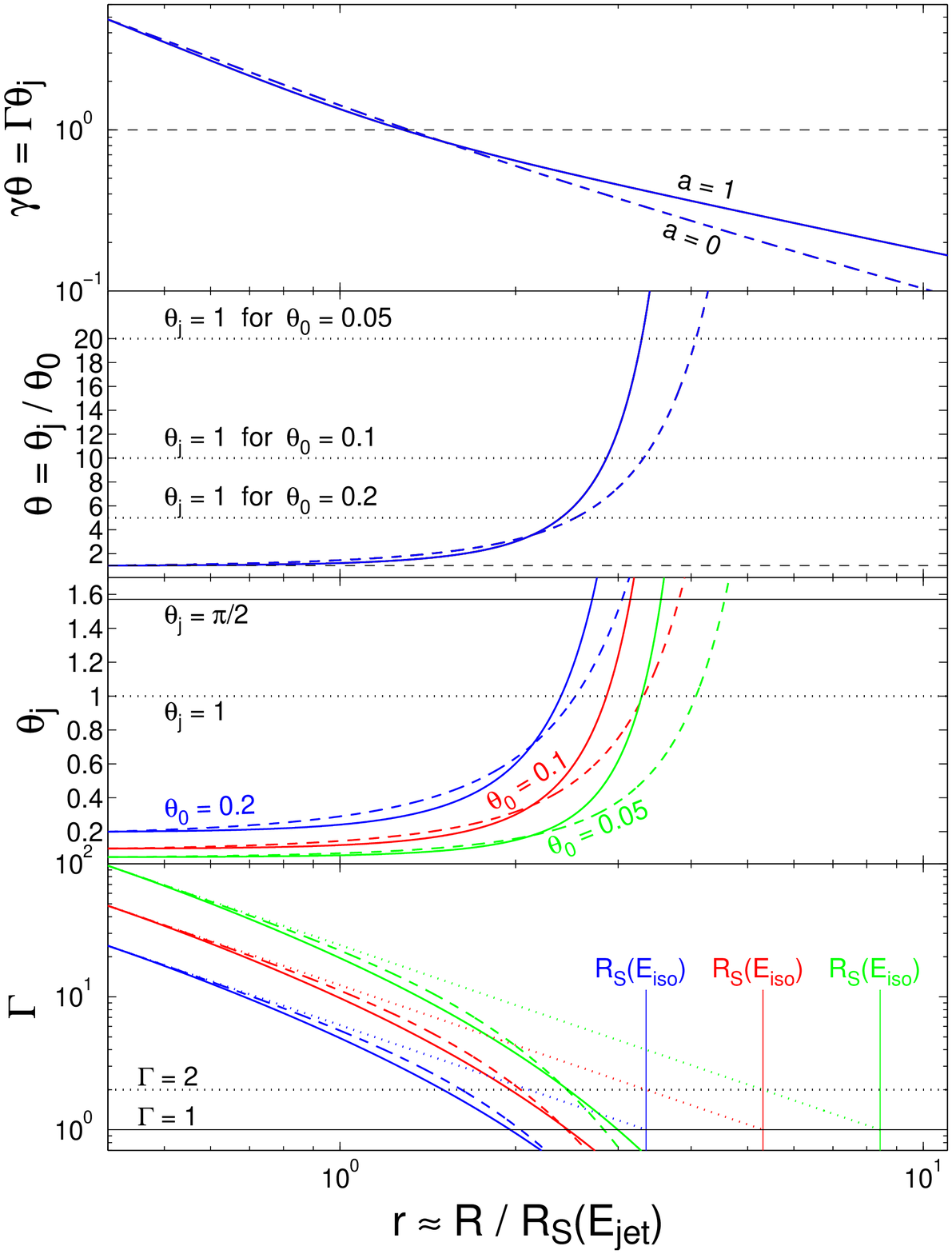}
\caption{Similar to Fig.~\ref{fig:jet_dyn_1D} but only for a uniform 
external density ($k=0$) and for three different values of the initial
jet half-opening angle: $\theta_0 = 0.05$ ({\it green}), $\theta_0 =
0.1$ ({\it red}) and $\theta_0 = 0.2$ ({\it blue}). The old ($a=0$)
and new ($a=1$) recipes for the jet lateral expansion are sown by {\it
dashed} and {\it solid} lines, respectively. In the top two panels the
lines for different $\theta_0$ and the same $a$ values coincide (see
text for details).  The values of $R_{\rm S}(E_{\rm iso})$ are
indicated in the bottom panel for reference.
\label{fig:jet_dyn_1D_2}}
\end{center}
\end{figure}

\begin{figure}
\begin{center}
\includegraphics[width=0.85\columnwidth]{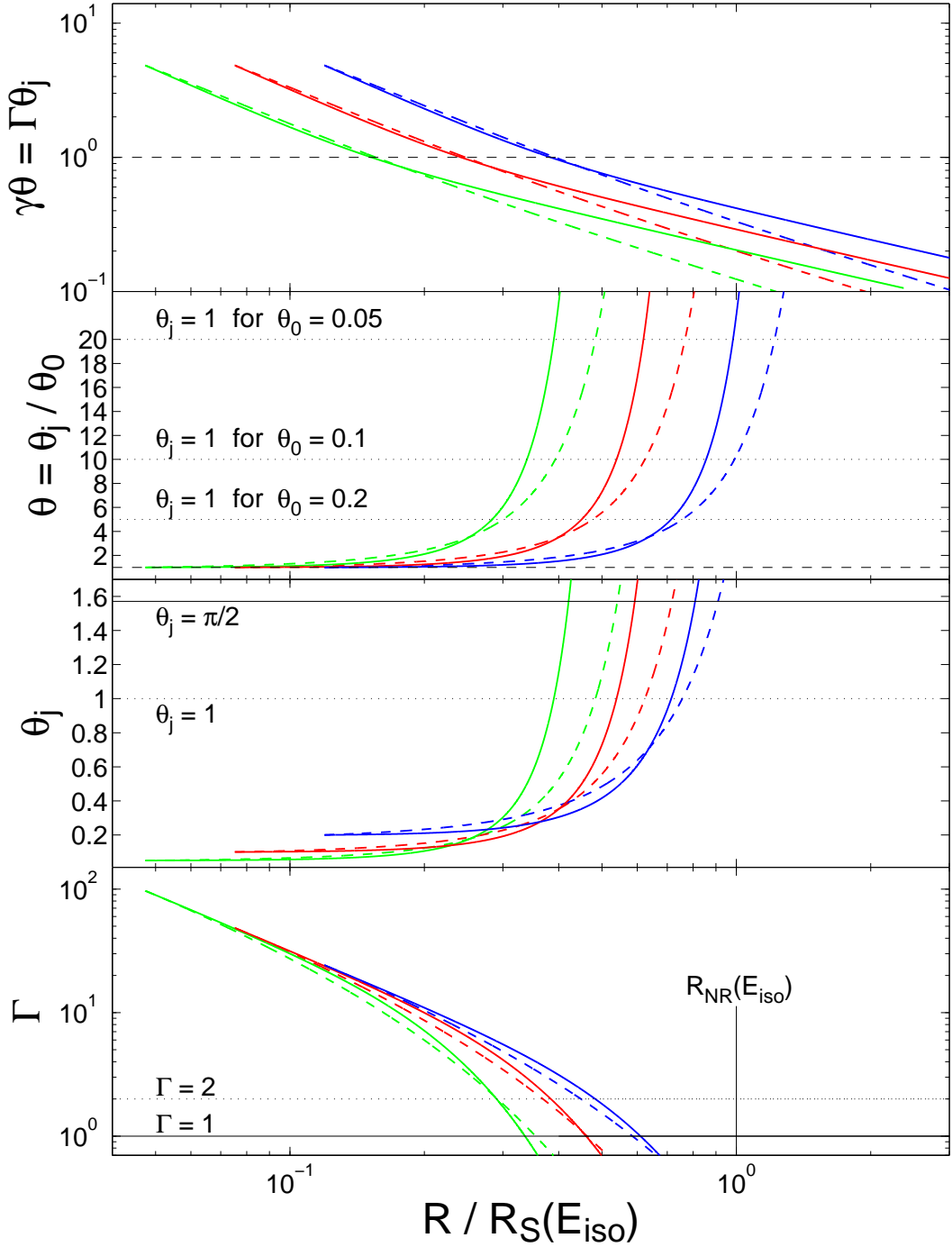}
\caption{Similar to Fig.~\ref{fig:jet_dyn_1D_2} but shown as a 
function of the jet radius $R$ normalized by $R_{\rm S}(E_{\rm iso})=
R_j\theta_0^{-2/(3-k)}$ instead of of $[(3-k)/2]^{-1/(3-k)}R_j =
[(3-k)/4]^{-1/(3-k)}R_{\rm S}(E_{\rm jet})$, where $E_{\rm iso}
\approx E_{\rm jet}2/\theta_0^2$ is the isotropic equivalent energy in
the jet, while $E_{\rm jet}$ is its true energy.
\label{fig:jet_dyn_1D_3}}
\end{center}
\end{figure}

Figure~\ref{fig:jet_dyn_1D_2} shows similar results for a uniform
external medium ($k=0$) and for three different values of the initial
jet half-opening angle, $\theta_0 = 0.05,\,0.1,\,0.2$. Since the
dynamical equations (Eqs.~[\ref{dtheta_dRt}] and [\ref{dGamma_dRt}])
involve only the normalized variables $\theta$, $\gamma$ and $r$, and
the initial conditions (Eq.~[\ref{IC}]) for $\theta$ and $\gamma$
depend only on the initial normalized radius $r_0$, the lines for
these normalized variables in the top two panels for the different
$\theta_0$ values exactly coincide.\footnote{This is since the same
value of $r_0 = 0.4$ was used, but in the limit $r_0\ll 1$ the
dependence of the solution on $r_0$ goes away at $r\gg r_0$.} The two
bottom panels show the un-normalized quantities $\theta_j$ and
$\Gamma$ for our three values of $\theta_0$. In the bottom panel we
have added for comparison the Sedov radius, $R_{\rm S}(E_{\rm iso})$,
for a spherical flow with the same isotropic equivalent energy the jet
started with. We define $R_{\rm NR}$ for our model as the radius where
formally $\Gamma = 1$ (at which point this model clearly breaks
down). Figure~\ref{fig:jet_dyn_1D_3} is similar to
Fig.~\ref{fig:jet_dyn_1D_2} but the jet radius $R$ is normalized by the
radius $R_{\rm S}(E_{\rm iso}) = \theta_0^{-2/(3-k)}R_j$ instead of
$[(3-k)/2]^{-1/(3-k)}R_j = [(3-k)/4]^{-1/(3-k)}R_{\rm S}(E_{\rm
jet})$.

\begin{figure}
\begin{center}
\includegraphics[width=0.82\columnwidth]{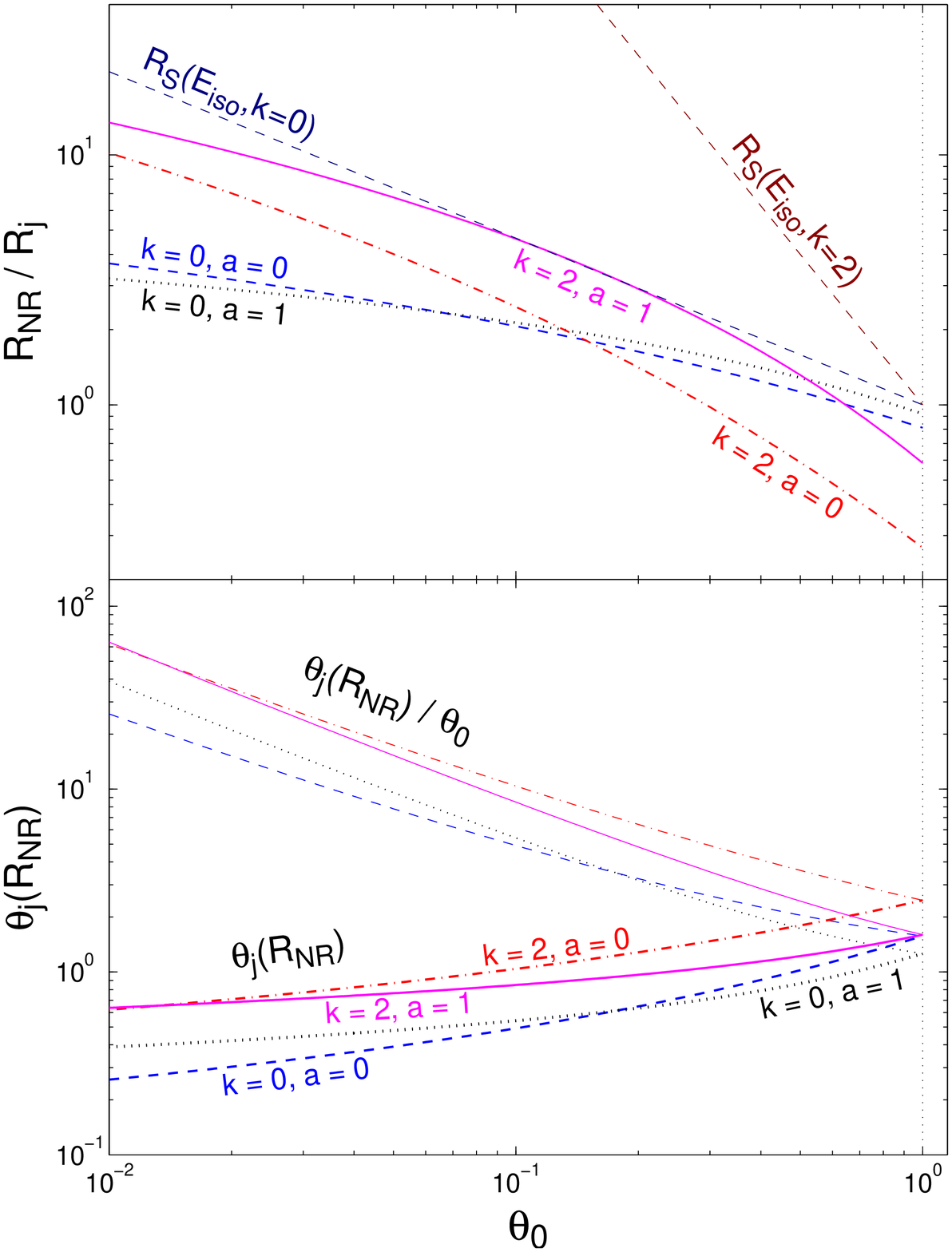}
\caption{{\bf Upper panel}: the non-relativistic transition radius for 
our analytic relativistic model, $R_{\rm NR}$, defined by
$\Gamma(R_{\rm NR}) = 1$, normalized by $R_j$, as a function of
$\theta_0$. For comparison, $R_{\rm S}(E_{\rm iso})$ is also shown;
the two radii, $R_{\rm NR}$ and $R_{\rm S}(E_{\rm iso})$, become
similar at $\theta_0\sim 1$ but are very different for $\theta_0\ll
1$. {\bf Lower panel}: the value of the jet half-opening angle,
$\theta_j$, at $R_{\rm NR}$, where our simple analytic relativistic
model breaks down.
\label{fig:theta_j_R_NR}}
\end{center}
\end{figure}

Fig.~\ref{fig:theta_j_R_NR} shows $R_{\rm NR}/R_j$ and
$\theta_j(R_{\rm NR})$ as a function of $\theta_0$. It can be seen
that $R_{\rm NR}$ depends on $\theta_0$ only logarithmically (as
can also be seen from Eq.~[\ref{Gamma_scaling}]), while $R_{\rm
S}(E_{\rm iso})/R_j = \theta_0^{-2/(3-k)}$ is simply a power of
$\theta_0$. It is also evident that $\theta_j(R_{\rm NR}) < 1$ for
$\theta_0\ll 1$, and its value increases with $\theta_0$ (while
$\theta_j(R_{\rm NR})/\theta_0$ decreases with $\theta_0$). This can
also be seen from Eq.~(\ref{y_j}), using the leading order term in $r$
and the definition $\Gamma(r_{\rm NR}) = 1$, which imply that
$\theta_j(r_{\rm NR}) = r_{\rm NR}^{-(3-k)/(3+a)}$, while $r_{\rm NR}$
(or $R_{\rm NR}$) decreases (logarithmically) with $\theta_0$. For $k
= 2$ the jet becomes non-relativistic and the model breaks down at
smaller values of $r = [(3-k)/2]^{1/(3-k)}R/R_j$ compared to $k = 0$,
which is consistent with the fact that the jet also starts to spread
sideways significantly at smaller values of $r$, of the order of
$r_c\approx\fracb{3-k}{2}^{(3+a)/[(1+a)(3-k)]}$. However, we are
primarily interested here in $k = 0$.

The model breaks down when $\Gamma$ drops to 1 (or even slightly
earlier). As can be seen from
Figs.~\ref{fig:jet_dyn_1D}--\ref{fig:jet_dyn_1D_3}, it breaks down
earlier for larger $\theta_0$ values, and its region of validity
(especially at $R\gtrsim R_j$) decreases as $\theta_0$ increases.  In
particular, for the value of $\theta_0 = 0.2$, which was most widely
used so far in numerical simulations~\citep[][while an even larger
value of $\theta_0 = 20^\circ \approx 0.35\;$rad was used in some
works -- \citealt{MK10};
\citealt{vanEerten11a}]{Granot01,ZM09,vanEerten10}, this
dynamical range is very narrow, and the asymptotic exponential growth
of $\theta_j$ with $R$ is not reached before the model breaks down (at
$\Gamma\lesssim 1.5-2$ or $\theta_j \gtrsim 0.5-1$). Even for
$\theta_0 = 0.05$, which was used in the most recent simulations
\citep{WWF11,vEM11} and is at the low end of the values inferred from
afterglow observations, the asymptotic exponential regime is only
barely reached before the model breaks down (in agreement with the
conclusions of \citealt{WWF11}). Note that in this (limited) region
of validity of this semi-analytic model our new recipe might still
result in smaller or comparable values of $\theta(r)$ (i.e. of
$\theta_j$ for a fixed $\theta_0$, at a given radius for a fixed
$E_{\rm jet}$) compared to the old recipe.  The discussion about when
this model breaks down is expanded in \S~\ref{sec:comp-sim}, where we
compare the analytic models to numerical simulations. Because of this
important limitation of our relativistic analytic model, in the next
section we generalize it so that it would not break down when the jet
becomes sub-relativistic or wide.

\section{Generalized models valid for arbitrary  $\Gamma$ and $\theta_j$}
\label{sec:gen}

In order to avoid the breakdown of the model at small Lorentz factors
$\Gamma$ or large jet half-opening angles $\theta_j$, we construct
here simple generalizations of the analytic model studied in the
previous section, which do not require the jet to be very narrow
($\theta_j\ll 1$) or highly relativistic ($u\approx\Gamma\gg 1$). Two
variants are introduced, named the trumpet model (in
\S~\ref{sec:trumpet}) and the conical model (in \S~\ref{sec:conical}),
according to the shape of the region from which the external medium is
assumed to have been swept up by the jet (before it becomes
spherical).

The rate at which the jet half-opening angle, $\theta_j$, increases
depends on the lateral velocity at the edge of the jet,
$\beta_\theta$, as $d\theta_j = \beta_\theta cdt/R =
(\beta_\theta/\beta_r)dR/R$, or
\begin{equation}
\frac{d\theta_j}{d\ln R} = \frac{\beta_\theta}{\beta_r}\ .
\end{equation}
A crude approximation for for the comoving 4-velocity of the lateral
expansion ($u'_\theta$), which would roughly correspond to the sound
speed both in the relativistic and in the Newtonian regimes, is
$u'_\theta \sim \beta = u(1+u^2)^{-1/2}$. This would modify the
traditional recipe to $\beta_\theta =
u'_\theta/\Gamma \sim \beta/\Gamma = u/(1+u^2)$ or $\beta_\theta/\beta_r
\sim \beta_\theta/\beta \sim 1/\Gamma = (1+u^2)^{-1/2}$. In our
recipe\footnote{Note that we use $\partial\ln u/\partial\theta =
u^{-1}\partial u/\partial\theta \sim -1/\Delta\theta$ since the
4-velocity $u$, unlike $\beta$ or $\Gamma$, generally varies
significantly with $\theta$ both in the relativistic and in the
Newtonian regimes, so that $\partial u/\partial\theta \sim
-u/\Delta\theta$ in both regimes, while $\partial\Gamma/\partial\theta
\sim -\Gamma/\Delta\theta$ only in the relativistic regime and
$\partial\beta/\partial\theta \sim -\beta/\Delta\theta$ only in the
Newtonian regime.} $\beta_\theta/\beta_r = -\partial\ln
R/\partial\theta \sim -\partial\ln\beta/\partial\theta
\sim -\Gamma^{-2}\partial\ln u/\partial\theta \sim 1/\Gamma^2\Delta\theta
\sim 1/\Gamma^2\theta_j = 1/[(1+u^2)\theta_j]$.  Therefore, just as
before, we still have
\begin{equation}\label{dtheta_dR_app}
\frac{d\theta_j}{d\ln R} = \frac{\beta_\theta}{\beta_r} 
\approx \frac{1}{\Gamma^{1+a}\theta_j^a}\ ,\quad\quad
a = \left\{\matrix{
1\quad\quad (\hat{\beta}=\hat{n})\ , \cr \cr
0\quad (u'_\theta\sim 1)\ .}\right.
\end{equation}

\subsection{The ``trumpet model''}
\label{sec:trumpet}

In this model we follow the usual assumption that the external rest
mass is swept-up by a working area consisting of the part of an
expanding sphere of radius $R$ within a half-opening angle
$\theta_j(R)$. Thus, the total swept-up (rest) mass, $M(R)$, for a
double-sided jet is accumulated as
\begin{equation}\label{dM_dR_app}
\frac{dM}{dR} \approx [1-\cos\theta_j(R)]4\pi R^2\rho_{\rm ext}(R) 
=  [1-\cos\theta_j(R)]4\pi AR^{2-k}\ .
\end{equation}
Energy conservation takes the approximate form $E_{\rm jet}\approx u^2 Mc^2$,
implying $Md(u^2) =-u^2dM$, and
\begin{equation}\label{dGamma_dR_app}
\frac{du}{dR} = -\frac{u}{2M}\frac{dM}{dR} =
-\frac{2\pi Ac^2}{E_{\rm jet}} R^{2-k}[1-\cos\theta_j(R)]u^3(R)\ .
\end{equation}
Thus, in terms of $r = [(3-k)/2]^{1/(3-k)}R/R_j$ we have
\begin{equation}\label{du_dRt}
\frac{d\theta_j}{d\ln r} \approx \frac{1}{(1+u^2)^{(1+a)/2}\theta_j^a}\ ,\quad\quad
\frac{du}{dr} = -r^{2-k}u^3(r)2[1-\cos\theta_j(r)]\ ,
\end{equation}
where the initial conditions at some small radius $R_0 \ll R_{\rm S}(E_{\rm jet})
\sim R_j$ (just after the deceleration radius), corresponding to $r_0$, are given by
\begin{equation}\label{IC_app}
\theta_j(r_0) = \theta_0\ ,\quad\quad
u(r_0) =
\sqrt{\frac{3-k}{4(1-\cos\theta_0)}}\,r_0^{-(3-k)/2}\ .
\end{equation}

\subsection{The ``conical model''}
\label{sec:conical}

Here we note that the usual assumption that leads to
Eq.~(\ref{dM_dR_app}) neglects the external matter at the sides of the
jet. Because of this, when eventually $\theta_j$ reaches $\pi/2$ at
$R_{\rm sph}$ and is thus assumed to be fully spherical, the amount of
swept-up external rest mass at $R_{\rm sph}$ calculated according to
Eq.~(\ref{dM_dR_app}) will be significantly smaller than that
originally within a sphere of the same radius. Therefore, here in the
conical model we adopt an alternative approach of using for the rest
mass of the swept-up matter, that originally within a cone of
half-opening angle $\theta_j$,
\begin{equation}\label{dM_dR_app2}
M(R) \approx [1-\cos\theta_j(R)]\frac{4\pi}{(3-k)}A R^{3-k}\ .
\end{equation}
This still has the drawback of assigning the same Lorentz factor to
all of the swept-up external matter, even though that at the sides of
the jet should have a significantly smaller 4-velocity than that near
the head of the jet.  Using a slightly different normalized radius,
$r_{\rm S} = R/R_{\rm S}(E_{\rm jet}) = 2^{1/(3-k)}R/R_j =
[4/(3-k)]^{1/(3-k)}r$, energy conservation ($E_{\rm jet}\approx u^2
Mc^2$) and Eq.~(\ref{dtheta_dR_app}) imply
\begin{equation}\label{dtheta_drbar}
u(r_{\rm S}) = \frac{r_{\rm S}^{-(3-k)/2}}{\sqrt{1-\cos\theta_j(r_{\rm S})}}\ ,\quad\quad
\frac{d\theta_j}{d\ln r_{\rm S}} \approx 
\frac{1}{[1+r_{\rm S}^{k-3}(1-\cos\theta_j)^{-1}]^{(1+a)/2}\theta_j^a}\ ,
\end{equation}
where the initial conditions at some small radius $R_0 \ll R_{\rm NR,sph}(E)
\sim R_j$, corresponding to $r_{\rm S,0}$, are given by
\begin{equation}\label{IC_app2}
\theta_j(r_{\rm S,0}) = \theta_0\ ,\quad\quad
u(r_{\rm S,0}) = \frac{r_{\rm S,0}^{-(3-k)/2}}{\sqrt{1-\cos\theta_0}}\ .
\end{equation}

\subsection{Results for the generalized models}
\label{sec:model-results}

\begin{figure}
\begin{center}
\includegraphics[width=0.85\columnwidth]{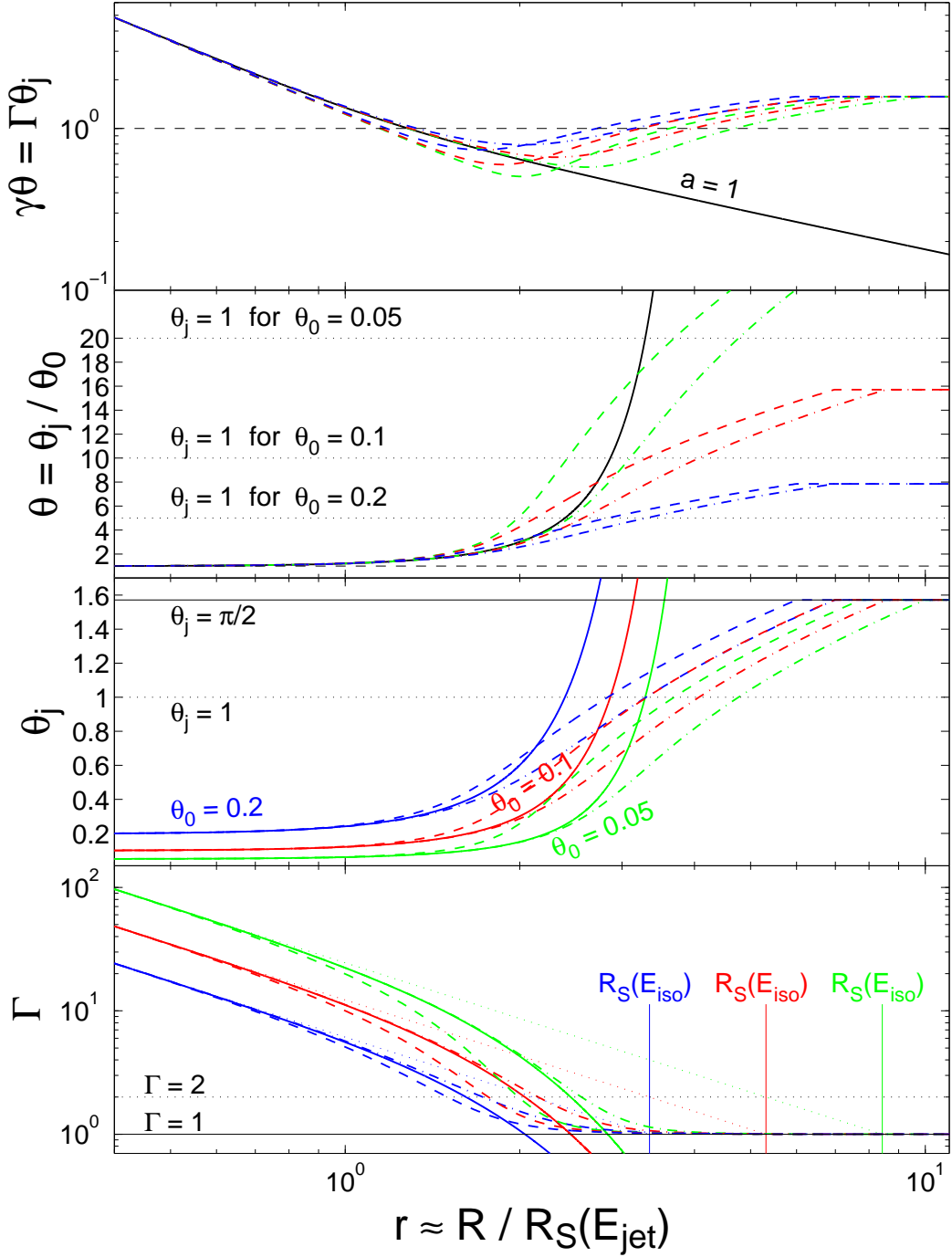}
\caption{Comparison between our relativistic ({\it solid lines}), 
trumpet ({\it dot-dashed lines}) and conical ({\it dashed lines})
models, where all models use our new recipe for the lateral spreading
of the jet ($a = 1$), and for a uniform external medium
($k=0$). Results are shown for three different values of the jet
initial half-opening angle: $\theta_0 = 0.05$ (in {\it green}),
$\theta_0 = 0.1$ (in {\it red}), and $\theta_0 = 0.2$ (in {\it
blue}). For reference we also indicate the values of $\Gamma\theta_j =
1$ in the top panel, some relevant values of $\theta_j$ in the two
middle panels, as well as the values of $R_{\rm S}(E_{\rm iso})$ and
$\Gamma = 1,\,2$ in the bottom panel.
\label{fig:jet_dyn_1D_2b}}
\end{center}
\end{figure}

\begin{figure}
\begin{center}
\includegraphics[width=0.85\columnwidth]{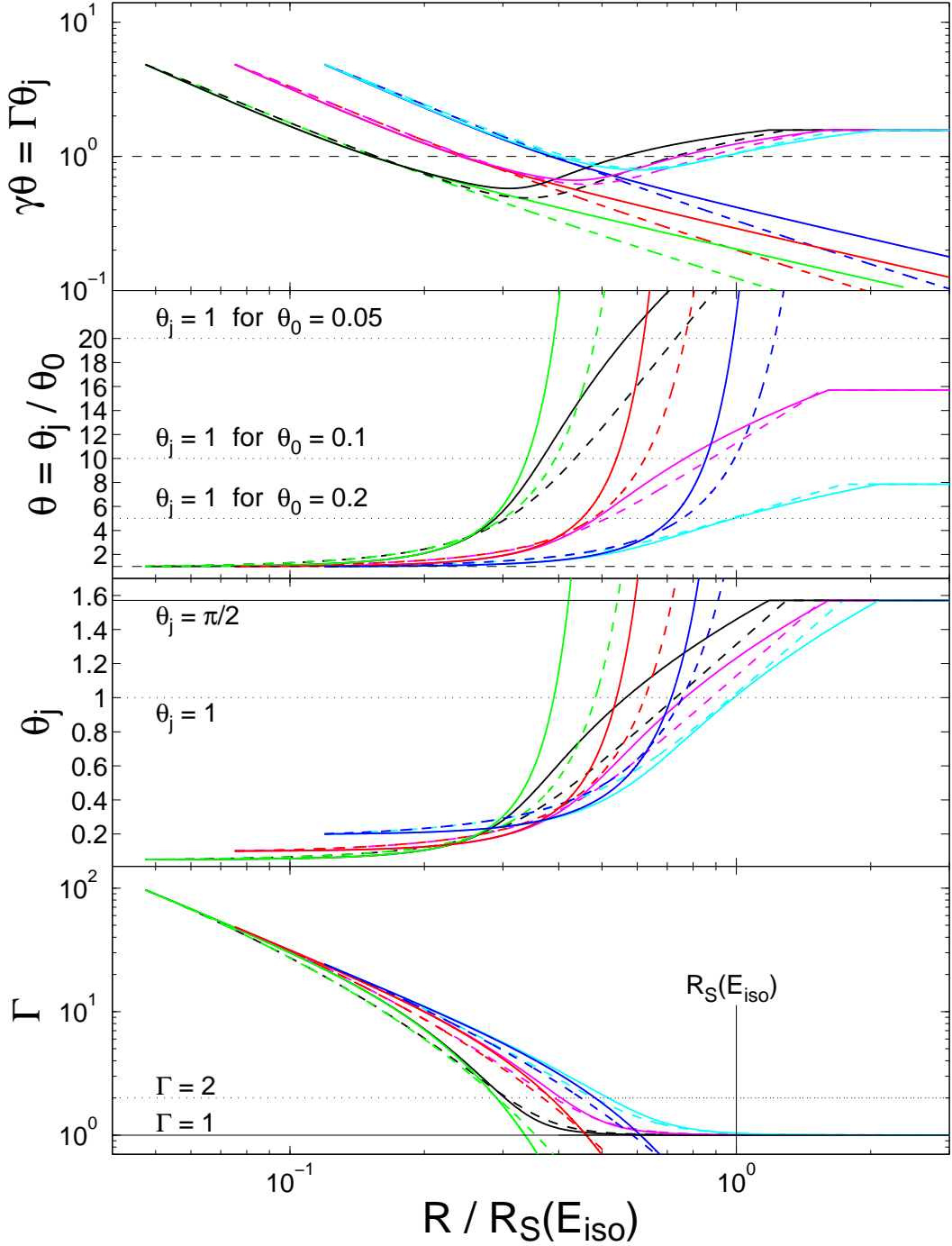}
\caption{Similar to Fig.~\ref{fig:jet_dyn_1D_2b} but shown (1) only 
for our relativistic model ({\it green}, {\it red}, and {\it blue}
lines for $\theta_0 = 0.05$, 0.1, and 0.2, respectively) and trumpet
model ({\it black}, {\it magenta}, and {\it cyan} lines for $\theta_0
= 0.05$, 0.1, and 0.2, respectively), (2) for both the old recipe ($a =
0$; {\it dashed lines}) and our new recipe ($a = 1$; {\it solid
lines}) for the jet lateral expansion, and (3) as a function of the jet
radius $R$ normalized by $R_{\rm S}(E_{\rm iso})=
R_j\theta_0^{-2/(3-k)}$ instead of of $[(3-k)/2]^{-1/(3-k)}R_j =
[(3-k)/4]^{-1/(3-k)}R_{\rm S}(E_{\rm jet})$.
\label{fig:jet_dyn_1D_3a}}
\end{center}
\end{figure} 

\begin{figure}
\begin{center}
\includegraphics[width=0.85\columnwidth]{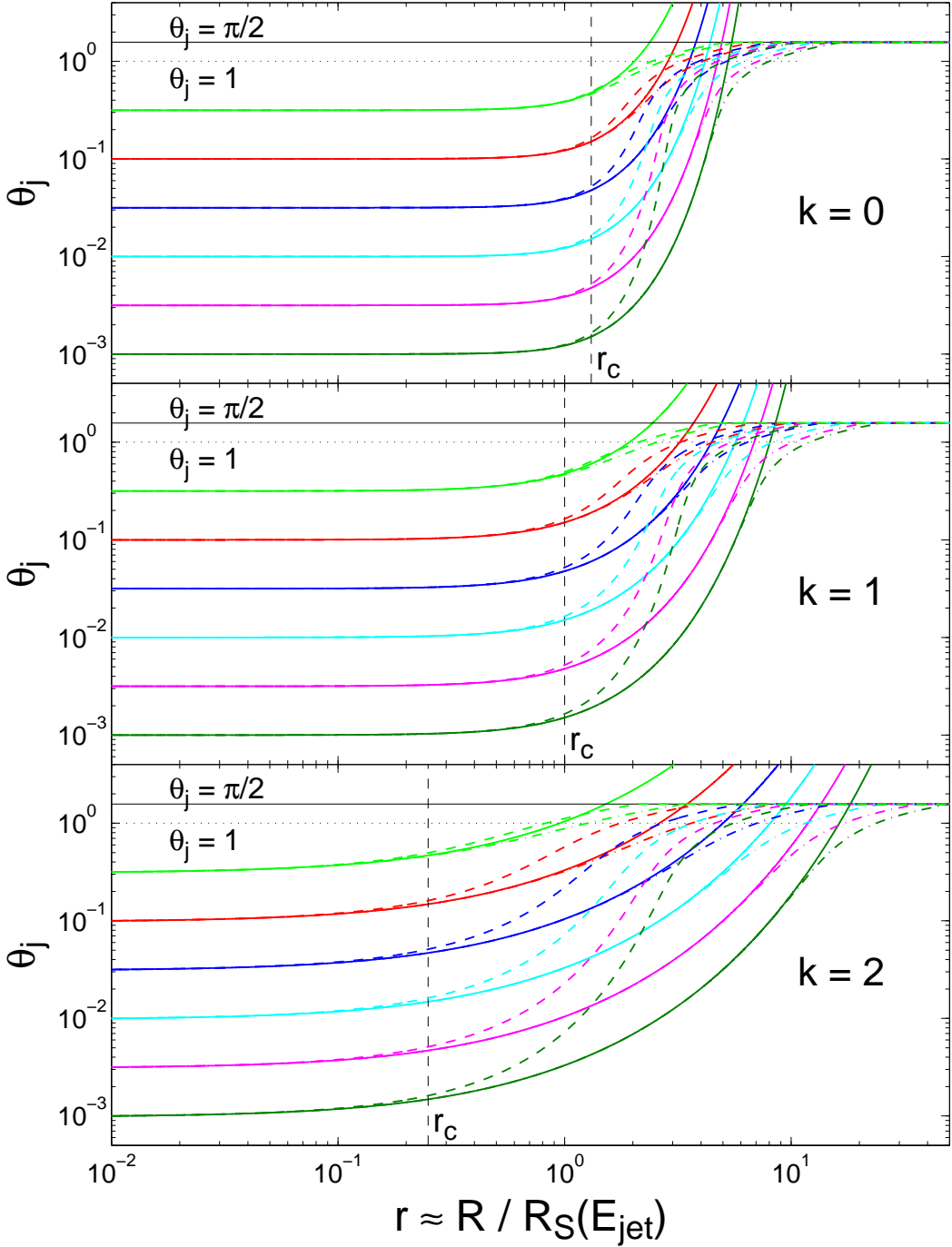}
\caption{Comparison between our relativistic ({\it solid lines}), 
trumpet ({\it dot-dashed lines}) and conical ({\it dashed lines})
models in terms of the evolution of the jet half-opening agle
$\theta_j$ with the normalized radius $r$, for $k = 0,\,1,\,2$ ({\it
top to bottom panels}), where all models use our new recipe for the
lateral spreading of the jet ($a = 1$).  Results are shown for
$\log_{10}(\theta_0) = -3,\,-2.5,\,...\,,\,-0.5$ ({\it using different
colors}) while the values of $\theta_0 = 1,\,\pi/2$ and the critical
radius $r_c$ (given by Eq.~[\ref{eq:r_c}], where the lateral spreading
is expected to become significant) are shown for reference.
\label{fig:jet_dyn_1D_new2c_k012}}
\end{center}
\end{figure}

\begin{figure}
\begin{center}
\includegraphics[width=0.85\columnwidth]{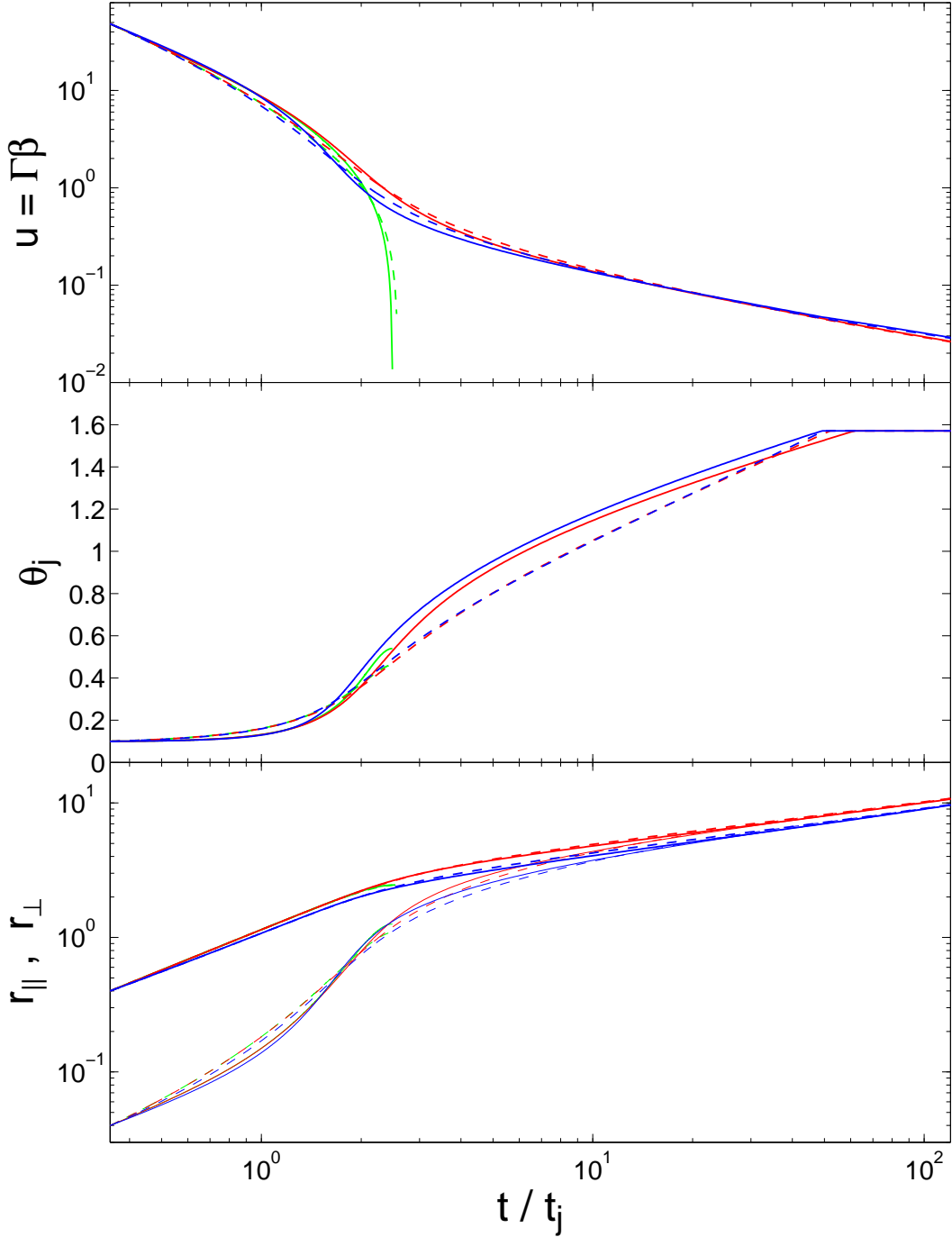} 
\caption{The jet dynamics according to our different analytic models, 
for $\theta_0 = 0.1$ and $k=0$. We show the jet 4-velocity, $u$
({\it upper panel}), half-opening angle, $\theta_j$ ({\it middle
panel}), as well as its normalized radius $r_\parallel = r$ and
lateral size $r_\perp = r\sin\theta_j$ ({\it bottom panel}), as a
function of the normalized lab frame time, $t/t_j$, for our
relativistic ({\it green lines}; until it breaks down at
$\Gamma\approx 1$), trumpet ({\it red lines}) and conical ({\it blue
lines}) models. The {\it solid} and {\it dashed} lines are,
respectively, for our new recipe ($a = 1$; Eq.~[\ref{eq:new-recipe}])
and the old recipe ($a=0$; Eq.~[\ref{eq:old-recipe}]) for the jet
lateral expansion.
\label{fig:R_T_lab4}}
\end{center}
\end{figure}

\begin{figure}
\begin{center}
\includegraphics[width=0.85\columnwidth]{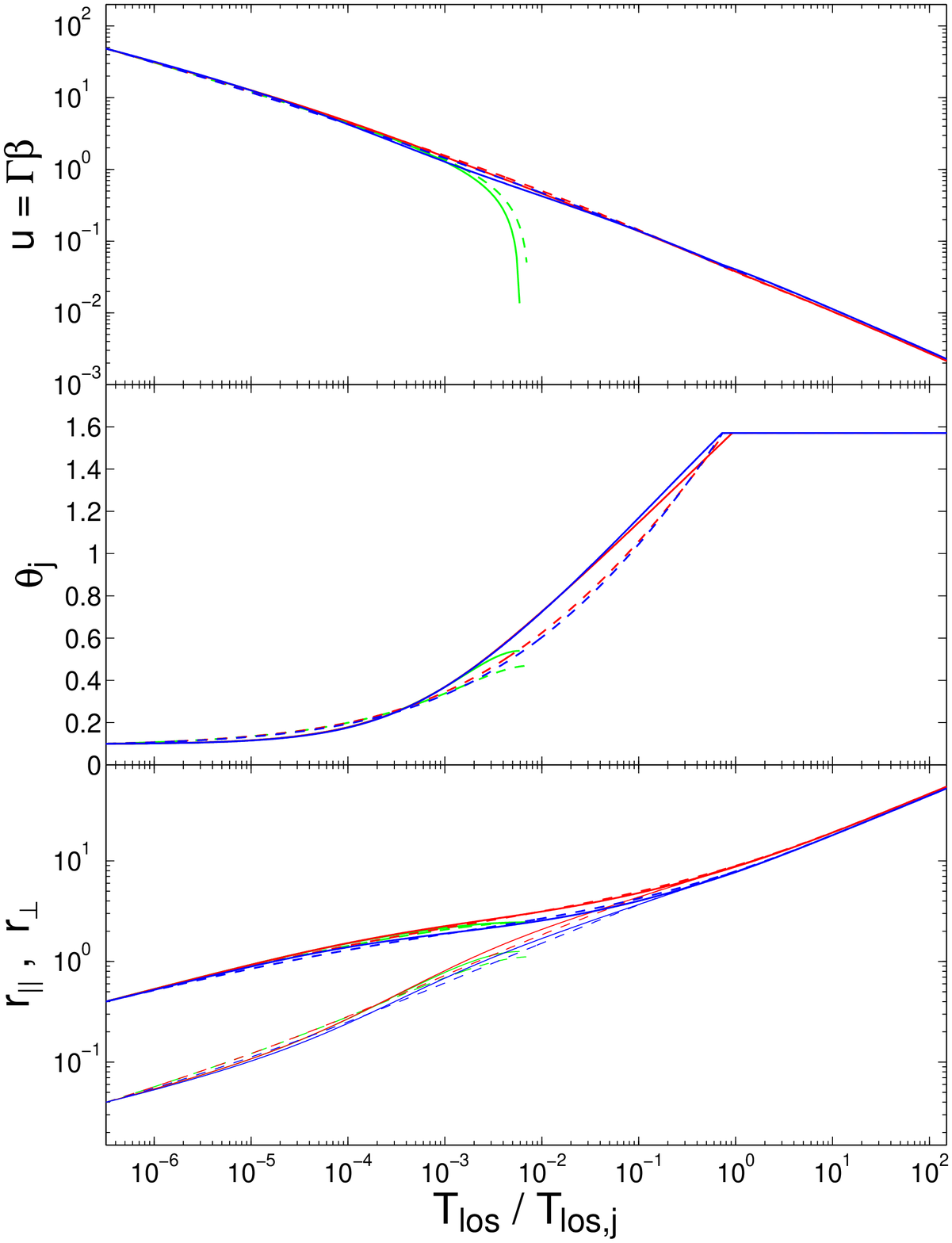}
\caption{Similar to Fig.~\ref{fig:R_T_lab4} but as a function of the
observed time, $T_{\rm los}$, at which photons from the front of the
jet reach an observer located along its symmetry axis, normalized by
its value at the jet break time $T_{\rm los,j}$.
\label{fig:R_T_los4}}
\end{center}
\end{figure}

Figures~\ref{fig:jet_dyn_1D_2b}, \ref{fig:jet_dyn_1D_3a} and
\ref{fig:jet_dyn_1D_new2c_k012} depict a comparison of these two
models with the relativistic model. All three models agree at early
times, while the jet is still highly relativistic, narrow and hardly
expanded sideways. The approximations of our relativistic model hold
well at this stage and the difference in the swept-up mass between the
trumpet and conical models is still very small. At later times,
however, the three models show a different behavior. The main effect
of the relaxation of the small $\theta$ and ultra-relativistic
approximations is that for typical values of $\theta_0 \gtrsim 0.05$
the region of exponential growth of $\theta_j$ with $R$ largely
disappears, and is replaced by a much slower, quasi-logarithmic
growth. This can most clearly be seen by comparing the results of the
relativistic model (from \S~\ref{sec:model}; {\it solid lines} in
Figs.~\ref{fig:jet_dyn_1D_2b} and \ref{fig:jet_dyn_1D_new2c_k012}, and
green, red or blue lines in Fig.~\ref{fig:jet_dyn_1D_3a}) and the
trumpet model (from \S~\ref{sec:trumpet}; {\it dot-dashed lines} in
Figs.~\ref{fig:jet_dyn_1D_2b} and \ref{fig:jet_dyn_1D_new2c_k012}, and
black, magenta or cyan lines in Fig.~\ref{fig:jet_dyn_1D_3a}). These
two models share the same assumption on the accumulation of the
swept-up external medium, and differ only by relaxing in the trumpet
model the requirements of $\Gamma\gg 1$ and $\theta_j\ll 1$. The
results of these two models are very close at early times while
$\Gamma\gg 1$, but diverge as $\Gamma$ becomes more modest and the
simple relativistic model reaches the exponential regime. This
can also be seen in Fig.~\ref{fig:jet_dyn_1D_new2c_k012} through the
fact that $\theta_j(r)$ for the two models start diverging when
$\theta_j$ becomes modest and the small angle approximation breaks
down.

The main difference between the trumpet and conical models is that for
the conical model the swept up mass at a given radius $R$ is larger
than for the trumpet model, resulting in a smaller $\Gamma$ and
therefore also a larger $\theta_j$, i.e. a faster evolution of
$\theta_j$ and $\Gamma$ with $R$. Since the larger swept-up mass comes
from the sides of the jet, it becomes important only once the jet
starts expanding sideways significantly, which occurs at $r_c$ (see
Eq.~[\ref{eq:r_c}]). This can be clearly seen in
Fig.~\ref{fig:jet_dyn_1D_new2c_k012}, where the {\it dot-dashed}
(or {\it solid}, which practically coincide at early times) and {\it
dashed} lines, for the trumpet (or relativistic) and conical models,
respectively, start diverging near $r_c$. Note that this remains valid
for all $k$-values, while $r_c$ decreases with
$k$. Fig.~\ref{fig:jet_dyn_1D_new2c_k012} also shows that for
sufficiently small values of $\theta_0$, roughly $\theta_0\ll 0.05$
for $k = 0$ and even somewhat smaller $\theta_0$ values for larger $k$
values, there is still a phase of quasi-exponential lateral expansion
for $r_c \lesssim r\lesssim r(\theta_j\sim 10^{-0.5})$ or
$1.5\theta_0\lesssim \theta_j\lesssim 10^{-0.5}$. For such extremely
small values of $\theta_0$ the difference between the conical and
trumpet models becomes large during the exponential sideways expansion
phase, where the lateral expansion is faster in the conical model. We
note, however, that such extremely narrow initial jet half-opening
angles are below the smallest values that have so far been reliably
inferred from GRB afterglow modeling, so that they might not be very
relevant in practice.

Figs.~\ref{fig:R_T_lab4} and \ref{fig:R_T_los4} show the jet dynamics
according to our different analytic models, for $\theta_0 = 0.1$.  It
can be seen the the differences between the various models are rather
small until the point where our relativistic model breaks down.  The
behaviour of the jet radius ($R = R_\parallel$) and lateral size
($R_\perp$) as a function of the lab frame time ($t$) shows a lot of
similarities to the analytic expectations (compare the {\it bottom
panel} of Fig.~\ref{fig:R_T_lab4} to Fig.~\ref{fig:R1}). This, again,
demonstrates that our new recipe for the lateral spreading of the jet
results in slower lateral expansion compared to the old recipe (and is
closer to assumption 2 of no lateral spreading -- {\it dashed red
lines} in Fig.~\ref{fig:R1}) at early times when $\Gamma>\theta_j$ but
faster lateral expansion at late times when $\Gamma<\theta_j$
(i.e. closer to assumption 1 of fast lateral spreading -- {\it solid
blue lines} in Fig.~\ref{fig:R1}).

\section{Comparison with numerical simulations}
\label{sec:comp-sim}

We turn now to a comparison of our analytic models with the results of
full 2D special relativistic hydrodynamic simulations. To do so one
needs first to define which quantities should be compared. This,
however, is not unique and can be done in different ways. For the
4-velocity, $u$, and as one (out of a few) reference value for the jet
half-opening angle, $\theta_j$, we use the weighted mean over the
energy $E$ in the lab frame (excluding rest energy) of $u$ and
$\theta$, respectively,
\begin{equation}
\mean{u}_E = \frac{\int dE\,u}{\int dE}\ ,\quad\quad
\mean{\theta}_E = \frac{\int dE\,\theta}{\int dE}\ .
\end{equation}
For the jet radius (or parallel size, $R_\parallel = R$) and lateral
size ($R_\perp$) we use:
\begin{equation}
\mean{R_\parallel} = \mean{z}_E = \frac{\int dE\, z}{\int dE}\ ,\quad\quad
\mean{R_\perp} = \mean{x}_E = \mean{y}_E = \frac{2}{\pi}\,\mean{r_{\rm cyl}}_E 
= \frac{2}{\pi}\,\frac{\int dE\, r_{\rm cyl}}{\int dE}\ .
\end{equation}
These averages reduce to  $R_\parallel = R_\perp$ (or
$\mean{R_\parallel} = \mean{R_\perp}$) for a spherical flow.

In order to perform a proper comparison to our analytic models, we
need to calculate similar averages for our jet, which at any given
time is the part of a thin spherical shell within a cone of
half-opening angle $\theta_j$. Thus, the radial integration drops out
and we are left only with an integral over $\mu$ between $\mu_j =
\cos\theta_j$ and 1,
\begin{equation}
\frac{R_\parallel}{R} = \frac{\int_{\mu_j}^1 d\mu\,\mu}{\int_{\mu_j}^1 d\mu}
= \frac{\sin^2\theta_j}{2(1-\cos\theta_j)}\ ,\quad\quad
\frac{R_\perp}{R} = 
\frac{2}{\pi}\,\frac{\int_{\mu_j}^1 d\mu\,\sqrt{1-\mu^2}}{\int_{\mu_j}^1 d\mu}
= \frac{2\theta_j - \sin(2\theta_j)}{2\pi(1-\cos\theta_j)}\ .
\end{equation}
We can see that $R_\parallel = R_\perp$ for $\theta_j = \pi/2$, as it should.

Similarly, one can calculate $\mean{\theta}_E$ as a proxy for
$\theta_j$ in our models,
\begin{equation}
\mean{\theta}_E = \frac{\int_{\mu_j}^1 d\mu\,\arccos(\mu)}{\int_{\mu_j}^1 d\mu}
= \frac{\int_0^{\theta_j}d\theta\,\theta\sin\theta}{1-\cos\theta_j}
= \frac{\sin\theta_j - \theta_j\cos\theta_j}{1-\cos\theta_j}\ .
\end{equation}
This shows that $\mean{\theta}_E \approx (2/3)\theta_j$ for
$\theta_j\ll 1$, while $\mean{\theta}_E = 1$ for $\theta_j = \pi/2$
(which is the value for any spherical flow, also one with a radial
profile) and $(2/3)\theta_j < \mean{\theta}_E < (2/\pi)\theta_j$ for
$0 < \theta_j < \pi/2$. One can also calculate the angle out to which
a fraction $f$ of the energy is contained (or the energy $100f$
percentile),
\begin{equation}
\theta_f = \arccos\left[1-f(1-\cos\theta_j)\right]\ ,
\end{equation}
and compare it to the corresponding value from the numerical
simulations.

\begin{figure}
\begin{center}
\includegraphics[width=0.85\columnwidth]{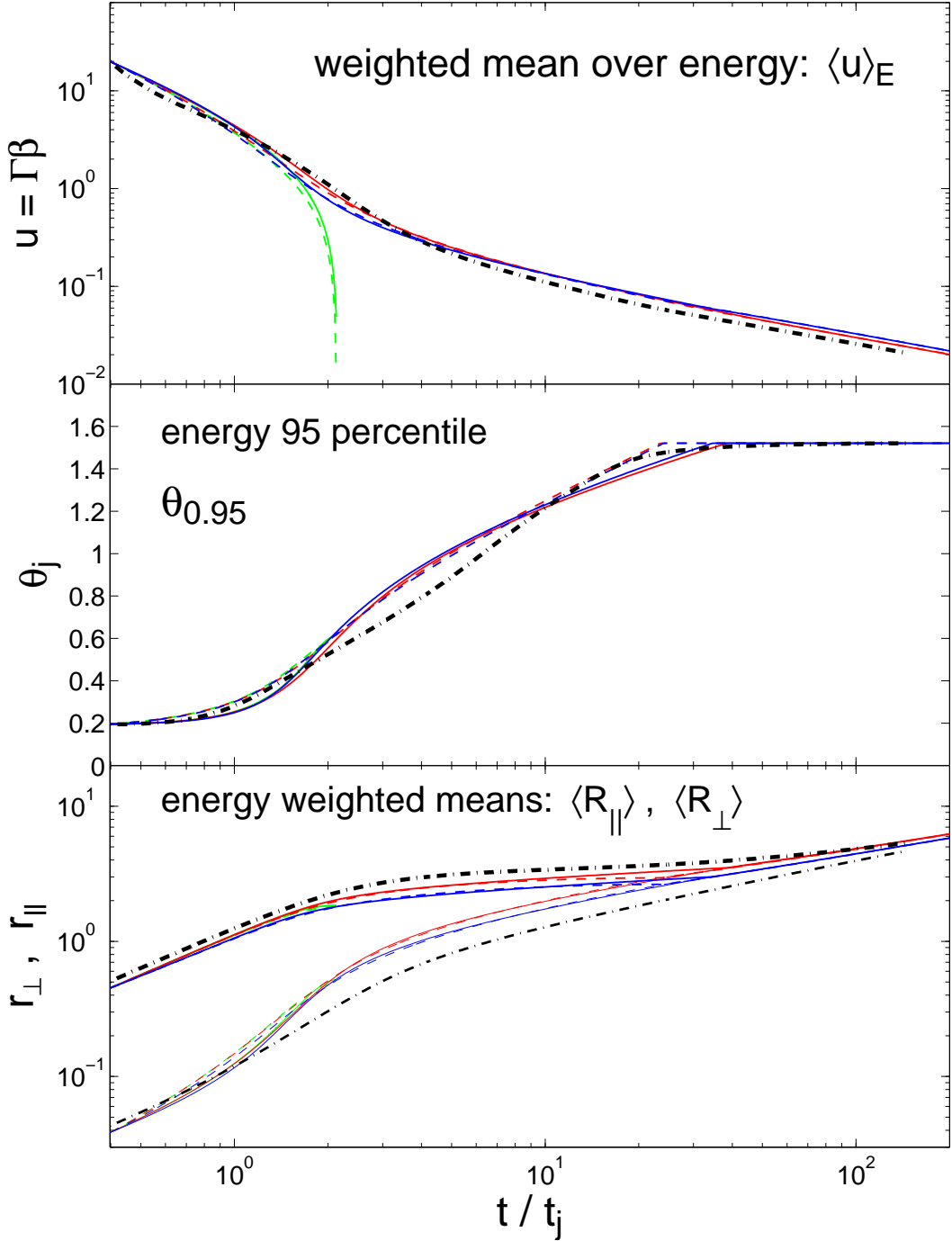}
\caption{Comparison, for $\theta_0 = 0.2$ and $k=0$, between our 
analytic models ({\it thin lines}) and the results of 2D special
relativistic hydrodynamic simulations
\citep[from][]{DeColle11a,DeColle11b} of a jet with initial conditions of
a conical wedge of half-opening angle $\theta_0$ taken out of the
\citet{BM76} self-similar solution ({\it thick dot-dashed black
line}), in terms of the jet 4-velocity ($u$), half-opening angle
($\theta_j$) as well as normalized parallel ($r_\parallel$) and
perpendicular ($r_\perp$) sizes. The {\it green}, {\it red} and {\it
blue} lines are for our relativistic, trumpet, and conical models,
respectively. Thin solid lines are for our new recipe for lateral
expansion ($a = 1$) while thin dashed lines are for the old recipe
($a=0$).
\label{fig:comp_sim2_k0}}
\end{center}
\end{figure}

\begin{figure}
\begin{center}
\includegraphics[width=0.85\columnwidth]{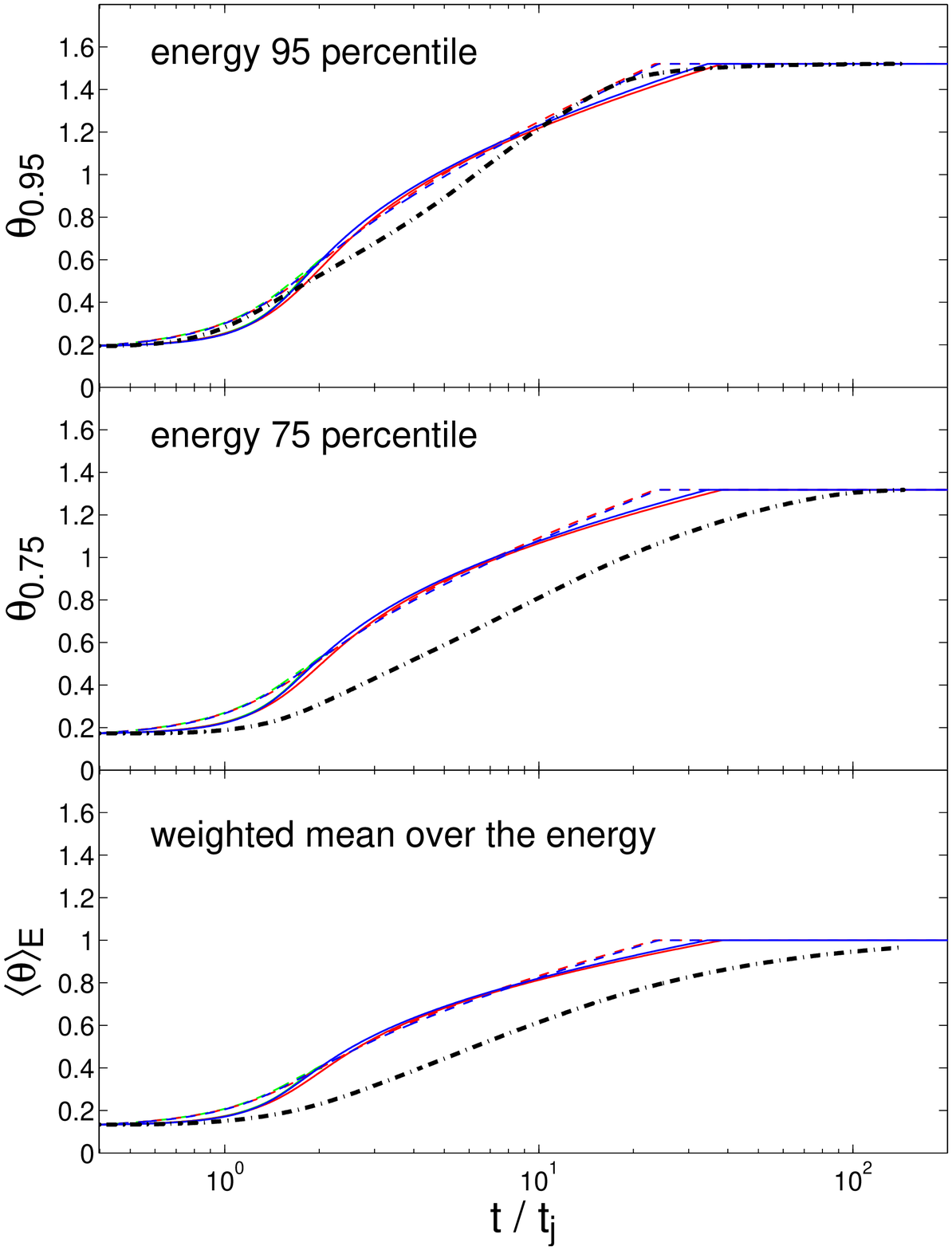}
\caption{Similar to Fig.~\ref{fig:comp_sim2_k0} but for three different
ways of quantifying the jet half-opening angle, $\theta_j$. The {\it
top panel} and {\it middle panel} show two different energy
percentiles, $\theta_{0.95}$ and $\theta_{0.75}$, respectively,
i.e. the values of $\theta$ up to which 95\% and 75\% of the energy is
contained.  The {\it bottom panel} shows the weighted mean over the
energy, $\mean{\theta}_E$.
\label{fig:comp_sim1a_k0}}
\end{center}
\end{figure} 

Figures~\ref{fig:comp_sim2_k0} and \ref{fig:comp_sim1a_k0} show a
comparison (for $k = 0$ and $\theta_0 = 0.2$) between the results of
our analytic models and of 2D special relativistic hydrodynamic
simulations~\citep[from][]{DeColle11a,DeColle11b}, when quantifying all
of them as discussed above. As can be seen from
Fig.~\ref{fig:comp_sim2_k0}, our models provide a reasonable overall
description of the full hydrodynamic simulations, and thus appear to
catch the basic underlying physics, despite their obvious simplicity.

Fig.~\ref{fig:comp_sim1a_k0} shows three different ways of quantifying
the jet half-opening angle, namely the weighted mean over the energy,
$\mean{\theta}_E$ ({\it bottom panel}), and two different energy
percentiles, $\theta_{0.75}$ ({\it middle panel}) and $\theta_{0.95}$
({\it top panel}), i.e. the values of $\theta$ up to which 75\% and
95\% of the energy, respectively, is contained.  It can be seen that
$\theta_{0.95}$ provides the best match between our analytic model and
the numerical simulations. For $\theta_{0.75}$ or $\mean{\theta}_E$
the match is not as good (though even then the difference is not very
large). This might be attributed to the fact that our analytic models
assume a uniform energy per solid angle, $\epsilon = dE/d\Omega$,
within the jet opening angle ($\theta <\theta_j$), while in practice
(or in the numerical simulations) it drops towards the outer edge of
the jet. The drop in $\epsilon$ from the jet axis towards its edge
causes both smaller values of $\mean{\theta}_E$ and smaller values of
$\theta_f$ for the lower energy percentiles (or $f$-values) relative
to a uniform jet with the same $\theta_f$ for a large energy
percentile (or $f$-value; e.g., $f = 0.95$ in our case). 
The results for our new recipe for the jet sideways expansion are
somewhat closer to the numerical simulations compared to the usual
recipe for $\mean{\theta}_E$ and $\theta_{0.75}$, while the usual
recipe is perhaps slightly closer for $\theta_{0.95}$. 

Both the analytic models and the numerical simulations show that the
flow becomes spherical more than a decade in time after it becomes
sub-relativistic (which may be quantified as the time when $\mean{u}_E
= 1$). This can be attributed to the fact that once the flow becomes
sub-relativistic its sound speed quickly drops, and so does the rate
of lateral expansion. Moreover, as the flow gradually becomes more
spherical the lateral gradients become smaller, which makes the flow
approach spherical symmetry more slowly.

The numerical simulations show that $\theta_f$ corresponding to lower
energy percentiles (or $f$-values) approach their asymptotic values
for a spherical flow at later times. This shows that the transfer of
energy to larger $\theta$-values is the slowest near the center of the
jet and larger near its edges, which may in turn be attributed to the
lateral gradient (say, of $\epsilon$) in the jet, which are smallest
near its center and largest near its edge.

\section{Discussion}
\label{sec:dis}

In this work we have introduced a new, physically motivated recipe for
the lateral expansion of the jet (in \S~\ref{sec:recipe}). It is based
on the jump conditions for oblique shocks of arbitrary 4-velocity,
which imply that the velocity of fluid just behind the shock front (in
the downstream region) is in the direction of the local shock normal
(i.e. perpendicular to the shock front at that location; $\hat{\beta}
= \hat{n}$, Eq.~[\ref{n_beta_hat}]) in the upstream rest frame (which
in our case is identified with the rest frame of the external medium
and the central source). Our new recipe for the lateral expansion rate
of the jet ($\beta_\theta \sim 1/\Gamma^2\theta_j$,
Eq.~[\ref{eq:new-recipe}]) has an extra factor of $\Gamma\theta_j$ in
the denominator relative to the usual recipe that has been used so far
($\beta_\theta\sim 1/\Gamma$, Eq.~[\ref{eq:old-recipe}]). This results
in slower lateral expansion relative to the usual (or old) recipe at
early times when $\Gamma > \theta_j$, but faster lateral expansion at
later times when $\Gamma<\theta_j$, i.e. once the lateral expansion
becomes significant.

Next (in \S~\ref{sec:model}), we have implemented our new recipe as
well as the old recipe in a simple analytic model for the jet
dynamics, which is valid only for high Lorentz factors ($\Gamma\gg 1$)
and narrow jet half-opening angles ($\theta_j\ll 1$). This model shows
an exponential lateral expansion for $\Gamma < \theta_0$, like previous
analytic models of this type. However, we demonstrate that for typical
values of the initial jet half-opening angle ($0.05 \lesssim
\theta_0\lesssim 0.2$) this model is valid only over a very limited
dynamical range for $\Gamma<\theta_0$, so that the asymptotic
exponential lateral expansion regime is hardly reached before the
model breaks down. This leads to a reasonable agreement with
numerical simulations over this limited range (as shown by
\citealt{WWF11} and in \S~\ref{sec:comp-sim}).

This motivated us (in \S~\ref{sec:gen}) to generalize our relativistic
model so that it would be valid for any values of $\Gamma$ and
$\theta_j$. This was done by switching to the 4-velocity $u$ (instead of
$\Gamma$) as the dynamical variable that we evolve (so that it would
vary significantly in both the relativistic and the Newtonian regimes),
and systematically not relying on any relativistic or small angle
approximations. Moreover, we have implemented two different assumptions
for the accumulation of the swept-up external rest mass, 
corresponding to a different variant of the model. The  trumpet model
makes the usual assumption that the working surface is the part of a
sphere of radius $R$ within a cone of half-opening angle $\theta_j(R)$ .
The conical model  assumes that all the mass within a
cone of half-opening angle $\theta_j(R)$ was swept-up, so that once
the flow becomes spherical the swept-up mass is equal to that
originally within a sphere of the same radius.

Our relativistic, trumpet and conical models all agree at early times
when the jet is still highly relativistic, narrow and hardly expanded
sideways ($\Gamma > \theta_0^{-1}\gg 1$). At this stage the
approximations of our relativistic model hold well and there are only
very small differences in the swept-up mass between the trumpet and
conical models. However, at later times when $\Gamma <\theta_0^{-1}$
the relativistic model enters a phase of rapid, exponential sideways
expansion and it quickly breaks down, before becoming spherical. We
note, however, that for a stratified or stellar wind like external
medium ($k=2$) the jet is closer to being spherical than for a uniform
or ISM like external medium ($k=0$; see {\it bottom panel} of
Fig.~\ref{fig:theta_j_R_NR}) when the relativistic model breaks down.

For the trumpet and conical models, which are valid for any $\Gamma$
or $\theta_j$, the phase of rapid, exponential sideways expansion
largely disappears for typical values of $\theta_0 \gtrsim 0.05$.
This occurs because the jet is no longer ultra-relativistic soon after
$\Gamma$ drops below $\theta_0^{-1}$, and once it becomes mildly or
sub-relativistic its sound speed and therefore its rate of lateral
expansion decrease compared to the ultra-relativistic regime.  The
conical model evolves somewhat faster than the trumpet model, since it
accumulates external rest mass also from the sides of the jet, and
thus it slows down faster. The smaller $\Gamma$
results in turn in even faster lateral expansion rate and a larger
$\theta_j$ (at a given radius $R$ or lab frame time $t$).

We compared (in \S~\ref{sec:comp-sim}) our analytic models to the
results of 2D special relativistic hydrodynamic
simulations~\citep[from][]{DeColle11a,DeColle11b}, finding that they
provide a reasonable description of the numerical results at all
times. Therefore, they can be used for analytic calculations of the
afterglow emission, and would provide more realistic results compared
to previous analytic models. The main factor that significantly
improves the agreement with simulations compared to previous analytic
models is the fact that we have generalized the model to be valid also
at modest Lorentz factors $\Gamma$ and large jet half-opening angles
$\theta_j$. Both our analytic generalized (trumpet and conical) models
and the numerical simulations show that the jet first becomes
sub-relativistic and only then gradually approaches spherical symmetry
over a long time. 

For typical initial half-opening angles ($\theta_0\gtrsim 0.05$) the
phase of rapid exponential lateral spreading is largely eliminated and
it is replaced by a quasi-logarithmic increase in $\theta_j$ with
radius $R$ or lab frame time $t$. \citet{vEM11} have stressed that
while noticeable sideways expansion starts for $\Gamma<\theta_0^{-1}$,
this initially involves only a small fraction of the total jet energy
in its outer parts, and the central parts of the jet that carry most
of its energy take longer to start spreading their energy to wider
angles. While it is true that the jet does not remain uniform, the
differences in the early growth of the angles $\theta_f$ containing
different fractions $f$ of the jet energy, normalized by their initial
value, is not very large -- less than a factor of 2 in lab frame time
or radius between $f = 0.95$ and $f = 0.5$, and tend to become smaller
for narrower $\theta_0$. This can be seen from Figure 4 of
\citet{vanEerten11b}, which also shows that as $\theta_0$ is gradually
decreased down to $0.05$, its initial growth becomes steeper and it
looks as if an early phase of exponential growth starts to develop,
contrary to what is claimed in
\citet{vEM11}. Therefore, we conclude that (i) although the uniform jet 
approximation used in our analytic models is obvious rather crude, it
nonetheless provides a reasonable description of the energetically
dominant part of the jet, and (ii) the prediction of our analytic
models that an early exponential sideways expansion phase should exist
for sufficiently small $\theta_0$ is not only consistent with the
existing simulations, but these simulations even show a hint for the
development of such a phase. This should obviously be tested more
thoroughly by simulations that reach even lower values of $\theta_0$.

A phase of exponential lateral spreading was first found
by~\citet{Rhoads99} and \citet{Piran00} using a simple analytic model.
Later,~\citet{Gruzinov07} found a self-similar solution with a similar
scaling. Our main conclusion (which is in agreement with
\citealt{WWF11}) is that such a phase will occur in practice only for
jets that are initially extremely narrow (with $\theta_0\ll 0.05$ or
so), while for more modest values of $\theta_0\gtrsim 0.05$ that are
more typically inferred in GRB jets, such a phase effectively does not
exist. This basically reconciles the long lasting apparent discrepancy
between analytic models and numerical simulations.

\acknowledgements 
We thank Fabio De Colle for sharing the results of his numerical
simulations.  This research was supported by the ERC advanced research
grant ``GRBs''.

\appendix
\section{Appendix: Comparison to previous works}
\label{sec:comp-prev}

We compare here our formulation for the jet lateral expansion rate,
Eq.~(\ref{eq:new-recipe}), with earlier work.  This formula was first
derived within the context of GRBs by \cite{KG03}, who provided two
different derivations. The first follows our line of argument and is
based on the orthogonality of the shock front and the velocity of the
fluid just behind it in the rest frame of the fluid ahead of the shock
(Eq.~[\ref{n_beta_hat}]). The second derivation involves an analysis
of the dynamical equations integrated over the radial profile.

A result similar to Eqs.~(\ref{eq:alpha}) and (\ref{eq:new-recipe})
was also recently derived by \citet{Lyut11}, based on an earlier work
by~\citet{Shapiro79}. \citet{Lyut11} has argued that it implies a
negligible lateral expansion as long as $\Gamma\gg 1$ unless
$\Delta\theta < 1/\Gamma^2$, suggesting that with this model one
obtains a slow sideways expansion, as seen in the numerical
simulations.  However, we note that the condition $\Delta\theta <
1/\Gamma^2$ corresponds to $\beta_\theta\sim 1$. This requirement is
too extreme since $\beta_\theta\sim 1$ would result in a
quasi-spherical flow within a single dynamical time (since in that
case $\beta_\theta\gtrsim\beta_r$). As is well known (see also
\S~\ref{sec:model}), the traditional recipe for lateral 
expansion (Eq.~[\ref{eq:old-recipe}]), namely $\beta_\theta \sim
1/\Gamma$, already gives an asymptotic exponential growth of
$\theta_j$ with $R$ (i.e. very rapid lateral expansion).

The earlier work by~\citet{Shapiro79} discusses two possible
approximations for the dynamics of a non-spherical relativistic blast
wave, both based on a thin shell approximation for the layer of
shocked external medium that carries most of the energy, but with
different additional assumptions: (i) the quasi-radial approximation
\citep[used in the Newtonian regime by][]{LP69} in which each part of
the shock is assumed to move in a radial trajectory as if it were part
of a spherical flow with the same local conditions (and in particular
the same energy per solid angle, excluding rest energy, $\epsilon =
dE/d\Omega$), and (ii) the~\citet{Komp60} approximation, that the
pressure behind the shock is uniform, i.e. the same at all locations
behind the shock and is proportional to the average energy density in
the region bounded by the shock front. The first approximation assumes that
the energy per solid angle in the flow (excluding rest energy) does
not change and remains equal to its initial value, $\epsilon(t,\theta)
= \epsilon(t_0,\theta)$. In this sense, it basically assumes no
lateral expansion (as the jet retains its initial angular structure in
$\epsilon(\theta)$ indefinitely), so that this is a model assumption
in this case rather than a result.

The second approximation, which was originally used by \citet{Komp60}
in the Newtonian regime, does not appear to be very appropriate for
the relativistic regime where the angular size of causally connected
regions is $\sim 1/\Gamma\ll 1$, so that that the local dynamics of a
small portion of the flow should not be affected by the average energy
per unit volume in the whole flow, which may be dominated by regions
that are not in causal contact with it. A simple example of how
the~\citet{Komp60} approximation violates causality in the
relativistic regime is that for a uniform external medium it implies
that the velocity of the shock front is uniform~\citep[i.e. depends
only on the lab frame time, but not on the location along the shock
front;][]{Shapiro79}, which necessarily implies that the flow must
approach spherical symmetry within a few dynamical times.\footnote{The
direction of the velocity of the fluid just behind the shock, which is
along the shock normal, might be initially non-radial, but since the
shock velocity is the same everywhere and highly relativistic, it
quickly approaches spherical symmetry, similar to the wave left by a
stone thrown into water, where the velocity of the surface water wave
is uniform and the wave front quickly forgets the shape of the stone
and becomes circular as its radius becomes larger than that of the
stone. In our case, within a few dynamical times $\epsilon$ becomes
essentially independent of $\theta$, since its local value is
dominated by the recently shocked material, where the shock Lorentz
factor is uniform.} This obviously violates causality, since as we
discussed in the introduction, a roughly uniform jet with reasonable
sharp edges cannot expand sideways significantly as long as
$\Gamma\theta_0\gg 1$, from causal considerations (since its bulk is
not in causal contact with its edges, and it does not ``know'' that it
is not part of a spherical flow and should thus start expanding
sideways).

\citet{Shapiro79} reaches the conclusion that the two
approximations give the same result in the extreme relativistic limit
only because he explicitly assumed that in the quasi-radial
approximation the energy per solid angle, $\epsilon = dE/d\Omega$, is
not only independent of time, but also independent of the location
along the shock front (this can be seen from the fact that his energy
integral is independent of $\theta$). This assumption quickly leads to
a quasi-spherical flow for a spherical external density profile, and
the non-spherical solutions obtained by \citet{Shapiro79} arise since
he considered an exponential atmosphere, which is a highly
non-spherical external density profile. The problem of interest for
us, namely the dynamics of GRB jets during the afterglow phase,
involves a non-uniform initial distribution of the energy per solid
angle, $\epsilon(t_0,\theta)$, and in such a case the two
approximations are not equivalent in the extreme relativistic limit.
Therefore, we conclude that neither of these two approximations appears
to be appropriate for studying the dynamics or degree of lateral
spreading of GRB jets during the afterglow phase.

\end{document}